\documentclass[aps,prl,reprint,groupedaddress,showpacs,amsmath,amssymb,twocolumn,floatfix]{revtex4-1}
\usepackage{graphicx}
\usepackage{bm}
\usepackage{color}
\usepackage[normalem]{ulem}
\usepackage{amsmath}
\usepackage{tabularx}

\usepackage[colorlinks=true]{hyperref}  
\hypersetup{
    bookmarks=true,         
    unicode=false,          
    pdftoolbar=true,        
    pdfmenubar=true,        
    pdffitwindow=false,     
    pdfstartview={FitH},    
    pdftitle={Zero-field superconducting diode effect in small twist-angle trilayer graphene},    
    pdfauthor={Lin et al.},     
    pdfsubject={},   
    pdfcreator={},   
    pdfproducer={}, 
    pdfkeywords={} {} {}, 
    pdfnewwindow=true,      
    colorlinks=true,       
    linkcolor=magenta, 
    citecolor=blue,        
    filecolor=magenta,      
    urlcolor=blue           
} 
\usepackage{cleveref}

\newcommand{\Rxx}{$R_{xx}$}
\newcommand{\Rxy}{$R_{xy}$}

\newcommand{\Bpara}{$B_{\parallel}$}

\newcommand{\WSe}{WSe$_2$}

\usepackage{pifont}
\newcommand{\cmark}{\ding{51}}%
\newcommand{\xmark}{\ding{55}}%

\usepackage{bbm}
\newcommand{\pdagger}{{\phantom{\dagger}}}
\newcommand{\SIsubsec}[1]{\vspace{1em}\noindent\textbf{#1.}}
\newcommand{\tableref}[1]{Table~\ref{#1}}
\newcommand{\equref}[1]{Eq.~(\ref{#1})}

\newcommand{\figref}[1]{Fig.~\ref{#1}}
\renewcommand{\vec}[1]{\boldsymbol{#1}}

\begin{document}

\title{Zero-field superconducting diode effect in small-twist-angle trilayer graphene}

\author{Jiang-Xiazi Lin$^{1}$$^{\ast}$}
\author{Phum Siriviboon$^{1}$$^{\ast}$}
\author{Harley D. Scammell$^{2}$}
\author{Song Liu$^{3}$}
\author{Daniel Rhodes$^{3}$}
\author{K. Watanabe$^{4}$}
\author{T. Taniguchi$^{5}$}
\author{James Hone$^{3}$}
\author{Mathias S. Scheurer$^{6}$}
\author{J.I.A. Li$^{1}$}
\email{jia\_li@brown.edu}

\affiliation{$^{1}$Department of Physics, Brown University, Providence, RI 02912, USA}
\affiliation{$^{2}$School of Physics, the University of New South Wales, Sydney, NSW, 2052
Australia}
\affiliation{$^{3}$Department of Mechanical Engineering, Columbia University, New York, NY 10027, USA}
\affiliation{$^{4}$Research Center for Functional Materials, National Institute for Materials Science, 1-1 Namiki, Tsukuba 305-0044, Japan}
\affiliation{$^{5}$International Center for Materials Nanoarchitectonics,
National Institute for Materials Science,  1-1 Namiki, Tsukuba 305-0044, Japan}
\affiliation{$^{6}$ Institute for Theoretical Physics, University of Innsbruck, Innsbruck, A-6020, Austria}

\affiliation{$^{\ast}$These authors contributed equally in this work.}

\date{\today}

\maketitle

\textbf{The critical current of a superconductor can be different for opposite directions of current flow when both time-reversal and inversion symmetry are broken.
Such nonreciprocal behavior in superconducting transport, which creates a superconducting diode, has recently been demonstrated experimentally by breaking these symmetries with an applied magnetic field \cite{Ando2020diodes} or by construction of a magnetic tunnel junction \cite{Diez2021magnetic}.
Here we report an intrinsic superconducting diode effect which is present at zero external magnetic field in mirror symmetric twisted trilayer graphene (tTLG). 
Such nonreciprocal behavior, with sign that can be reversed through training with an out-of-plane magnetic field, provides direct evidence of the microscopic coexistence between superconductivity and time-reversal symmetry breaking. In addition to the magnetic-field trainability, we show that the zero-field diode effect can be controlled by varying carrier density or twist angle. In accordance with these experimental controls, a natural interpretation for the origin of the intrinsic diode effect is an imbalance in valley occupation of the underlying Fermi surface, which likely leads to finite-momentum Cooper pairing and nematicity in the superconducting phase.}

\begin{figure*}
\includegraphics[width=0.8\linewidth]{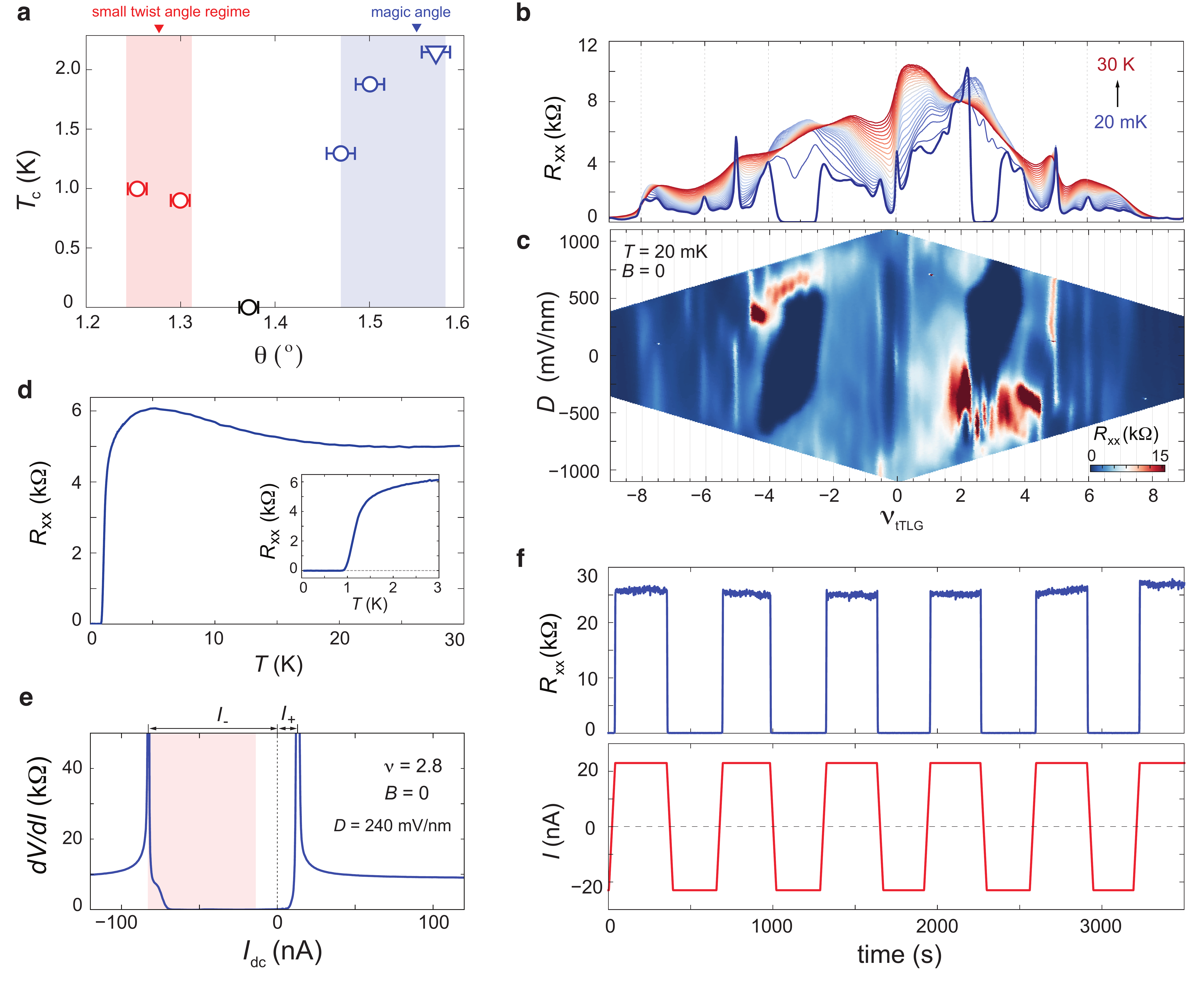}
\caption{\label{figSC}{\bf{Superconductivity and zero-field superconducting diode effect.}} (a) Twist-angle dependence of the superconducting transition temperature. The twist angle range near the magic angle is marked with the blue shaded area, whereas the small twist angle regime with red. The triangle is taken from Ref.~\cite{Park2021tTLG}. (b) Longitudinal resistance \Rxx\ as a function of moir\'e filling measured at $D = 0$, $B=0$ and different temperatures. (c) Longitudinal resistance \Rxx\ as a function of $D$ and moir\'e filling $\nu_{tTLG}$ measured at $B=0$ and $T= 20$ mK. 
(d) \Rxx\ as a function of temperature $T$ measured at $B=0$, $D=0$ and $\nu_{tTLG} = -3.29$. (e) Differential resistance as a function of DC current bias of the superconducting phase at $B=0$, $\nu_{tTLG}=2.59$ and $D = 400$ mV/nm. The measurement is performed after field-training with a positive magnetic field. $I_c^+$ and $I_c^-$ are the critical current with positive and negative DC current bias, respectively. (f) The zero-field superconducting diode effect is demonstrated by the switching between the normal and superconducting behavior as the applied DC current alternates between $\pm 22$ nA. The measurement is performed at $D=240$ mV/nm and $\nu_{tTLG} = 3.07$. } 

\end{figure*}

\begin{figure*}
\includegraphics[width=0.9\linewidth]{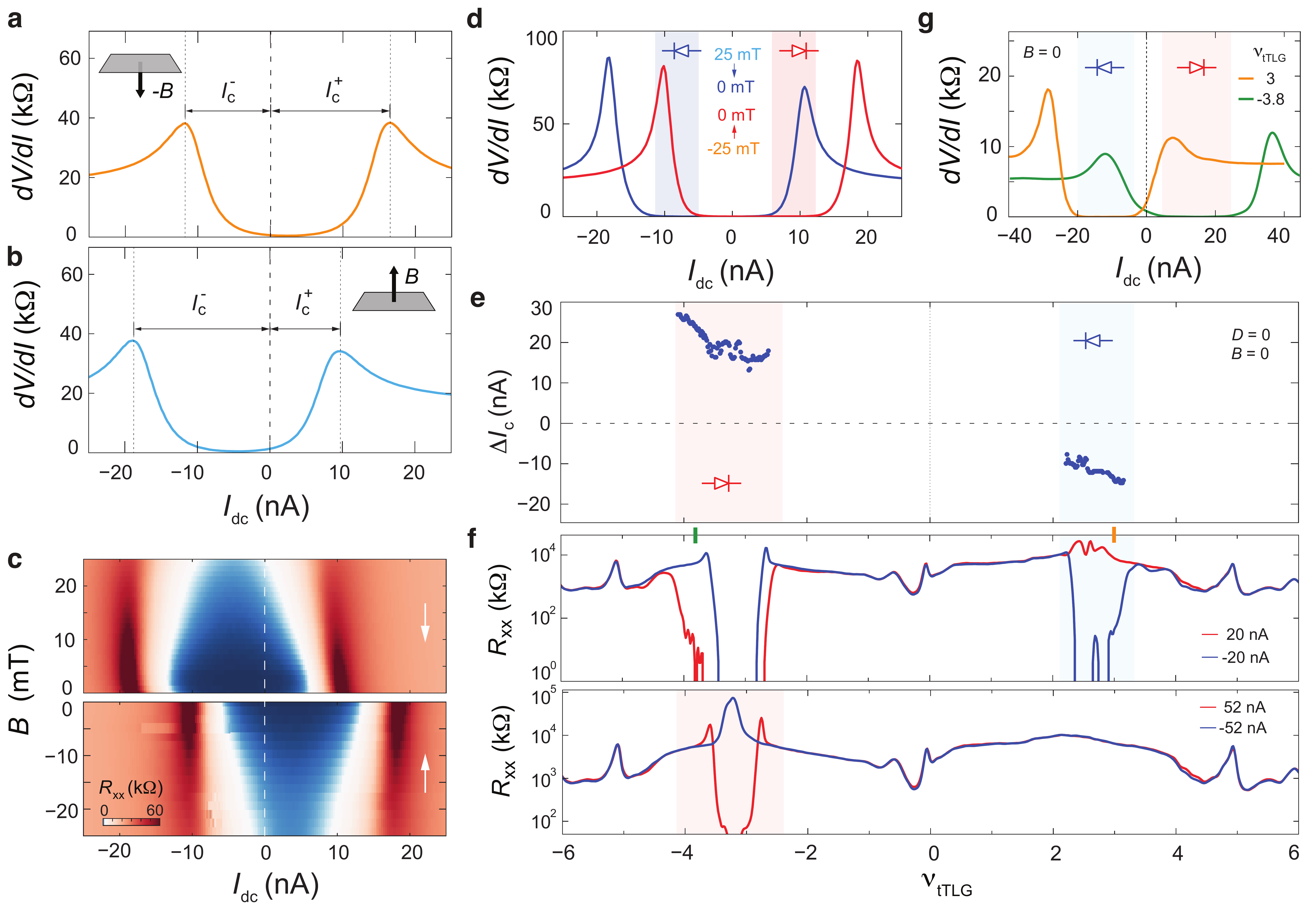}
\caption{\label{figdiode} {\bf{Controlling the superconducting diode effect.}}  (a-b) Differential resistance $dV/dI$ as a function of DC current $I_{dc}$ at $D = -573$ mV/nm, $\nu_{tTLG}=2.14$ and $T=20$ mK. The measurement is performed at (a) $B=-25$ mT and (b) $B=25$ mT. (c) $dV/dI - I_{dc} - B$ map. The vertical white arrow marks the direction in which the magnetic field is swept. (d) Differential resistance $dV/dI$ as a function of DC current $I_{dc}$ measured at $B=0$, $T=20$ mK, $D = -573$  mV/nm and $\nu_{tTLG}=2.14$. The blue and red traces in (d) are measured after field training at $B=+25$ mT and $-25$ mT, respectively. (e) The nonreciprocal component of the critical current, $\Delta I_c$, versus $\nu_{tTLG}$ measured at $B=0$ and $D=0$. This measurement is performed following a sequence of steps: (i) train the diode effect with a positive magnetic field; (ii) set the external magnetic field to zero; (iii) measure I-V characteristics at different carrier densities. Nonreciprocity is extracted from these I-V curves.  (f) \Rxx\ as a function of moir\'e filling measured at $D=0$, $B=0$ and $T= 20$ mK, after field-training with a positive magnetic field. The measurement is performed with an AC modulation of $1$ nA on top of a DC current bias of $\pm 20$ nA (top panel) and $\pm 52$ nA (bottom panel). (g) Differential resistance $dV/dI$ as a function of DC current $I_{dc}$ measured at $D = 0$, $T=20$ mK and different $\nu_{tTLG}$, as indicated.
}
\end{figure*} 

\begin{figure*}
\includegraphics[width=1\linewidth]{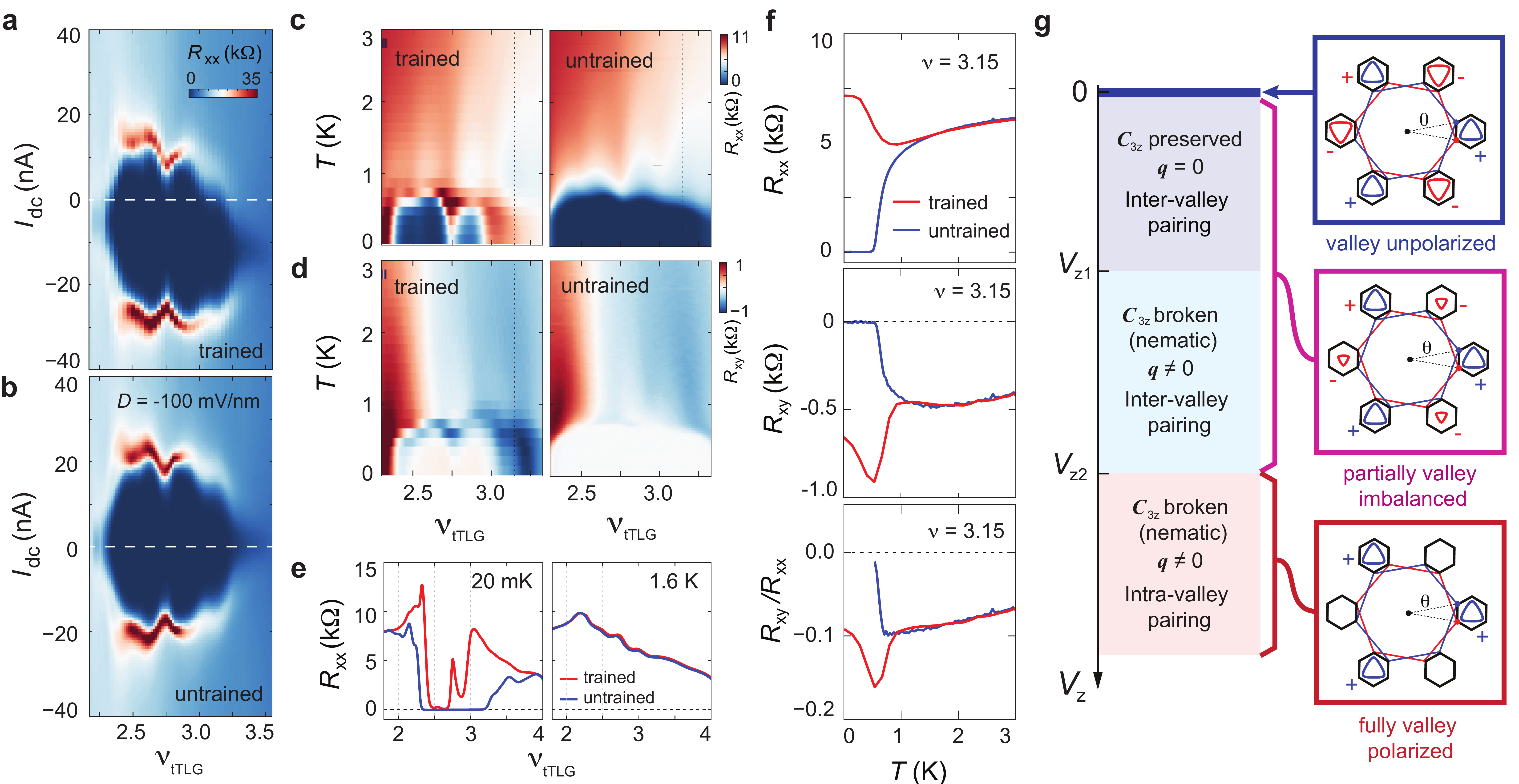}
\caption{\label{figT} {\bf{Possible origin of the zero-field superconducting diode effect.}} (a-b) Differential resistance as a function of moir\'e filling $\nu_{tTLG}$ and DC bias current measured at $T = 20$ mK and $B = 0$ mT. The measurement is performed (a) with the superconducting diode effect after field training, and (b) without the diode effect after ``un-training'' with a large DC current. (c) \Rxx\ and (d) \Rxy\ as a function of temperature $T$ and $\nu_{tTLG}$ measured with (left panels) and without (right panels) the superconducting diode effect.  The measurement is performed  with a small AC current bias of $I=1$ nA at $B=0$ and $D=0$ mV/nm.  
(e) \Rxx\ as a function of $\nu_{tTLG}$ with (red trace) and without (blue trace) the superconducting diode effect. The measurement is performed with a small AC current bias of $I=1$ nA at $B=0$, $D=0$, $T=20$ mK (left panel) and $T=1.6$ K (right panel). The superconducting diode effect induces dramatic changes in \Rxx\ at $T < T_c$ but has no influence at $T > T_c$. (f) Temperature dependence of (top panel) \Rxx, (middle panel) \Rxy, and (bottom panel) \Rxy$/$\Rxx, taken at $\nu_{tTLG}=3.15$, which is marked by vertical dashed lines in (b). (g) Schematic of the moir\'e Brillouin zone and Fermi surfaces in the presence of time-reversal symmetry. Valley polarization induces an imbalance between the Fermi surfaces in valley $+$ and $-$. The vertical axis on the left provides an overview of superconducting properties at different ranges of valley polarization $V_z$. $V_{z1}$ marks the critical valley polarization where Cooper pairs acquire non-zero center of mass momentum, $\vec{q} \neq 0$, and the Fermi surfaces are fully valley polarized for $V_z>V_{2z}$. For simplicity, we do not indicate that the superconducting state can be suppressed completely in a finite range of $V_z$ and that the onset of intra-valley pairing does generically not coincide with $V_{2z}$. 
}
\end{figure*}

As was noticed very early on in the field of superconductivity  \cite{Ginzburg1957ferromagnetic,Matthias1958spin,ANDERSON195926}, time-reversal symmetry, $\Theta$, is a key ingredient to the formation of a superconductor. This follows from the fact that, within BCS theory and for most known superconductors, Cooper pairs form between Kramers partners, whose degeneracy is the fundamental reason  why so many systems become superconductors at sufficiently low temperatures.
By the same token, breaking $\Theta$---by applying a magnetic field, as a consequence of magnetic impurities, or spontaneously at a magnetic instability---suppresses the superconducting state.  
Since the advent of unconventional superconductivity, the possibility that more exotic superconducting order parameters, beyond the BCS paradigm, can survive and coexist with  magnetic order has attracted significant attention to this day.
For example, sufficiently strong magnetic fields can give rise to the Fulde-Ferrell-Larkin-Ovchinnikov (FFLO) state \cite{Fulde1964FFLO,Larkin1965FFLO} where the center of mass momentum of Cooper pairs is non-zero, leading to a spatially modulated order parameter. 
Although the existence of non-uniform, FFLO-like states has been well-established by bringing a superconductor and a ferromagnet in close vicinity~\cite{Buzdin2005proximity}, and evidence of FFLO states in applied magnetic fields have been reported \cite{PhysRevLett.91.187004,FFLO2}, the microscopic coexistence between superconductivity and ferromagnetism remains rare~\cite{Coexistence1,Coexistence2,Coexistence3,Ren2009coexistence,Dikin2011coexistence,LuLi2011coexistence,Bert2011direct}.

Recently, an experimental study \cite{Ando2020diodes} demonstrated the non-trivial interplay of superconducting current and magnetic field, which attracted significant attention \cite{Yuan2021diodes,DaidoSCDiode,HeSCDiode}. It is shown that the magnetic-field-induced broken time-reversal symmetry together with the lack of inversion center in the system give rise to nonreciprocal supercurrents. This exotic phenomenon, often referred to as the superconducting diode effect, is manifested in an asymmetric current-voltage curve, where the superconducting critical current is different along opposite directions of DC current flow. Up to now, nonreciprocal transport behavior that is intrinsic to the superconductor has only been demonstrated in the presence of an external magnetic field ~\cite{Ando2020diodes,Ichikawa1994IV,Jiang1994IV,Broussard1988IV}. The realization of robust superconducting nonreciprocity at zero magnetic field would provide unambiguous evidence for the microscopic coexistence between superconductivity and spontaneous time-reversal symmetry breaking; not requiring an external field, it could further prove crucial for the construction of efficient non-dissipative quantum electronics.

In this work, we report the experimental observation of the zero-field superconducting diode effect in twisted trilayer graphene.
Fig.~\ref{figSC}a highlights two distinct twist angle regimes, which are marked with different colors. Out of 5 samples studied in this work (see Table~\ref{TableSample} for a list of samples), the zero-field superconducting diode effect is only observed in the small twist angle regime around $\theta = 1.3^{\circ}$, which is well below the expected magic angle. We note that an atomic interface between tTLG and a thin crystal of tungsten diselenide (\WSe) does not change the main phenomenology of our observation.  We will focus our discussion here on the behavior and tunability of the zero-field diode effect in sample A, which has a twist angle of $\theta = 1.25^{\circ}$, and leave the influence of the tTLG/\WSe\ interface to a separate discussion elsewhere.

In sample A, the robust superconducting phase emerges in the range of density $2 < \nu_{tTLG} < 4$ (Fig.~\ref{figSC}b). It occupies a portion of the phase space  in the $\nu_{tTLG}$-$D$ map which is shown as dark blue in the chosen color scale in Fig.~\ref{figSC}c. The temperature dependence of the longitudinal resistance \Rxx\ shows a sharp superconducting transition around $T = 1$ K, see Fig.~\ref{figSC}d. Notably, $T_c$ is not substantially suppressed compared to previous measurements near the magic angle, despite the fact that $\theta = 1.25^{\circ}$ is well below the magic angle.
Fig.~\ref{figSC}e plots the current-voltage characteristics measured at zero magnetic field, $B=0$, after field-training in a positive magnetic field. 

We see that the critical current of the superconducting phase, defined as the DC current bias at the onset of longitudinal resistance, is highly direction dependent. The nonreciprocal component of the critical current $\Delta I_c := I_c^{+}-I_c^{-}$, where $I_c^{+}$  and $I_c^{-}$ are the critical current with positive and negative DC bias current, respectively, is non-zero---the defining feature of the zero-field superconducting diode effect.
Since the critical current is much larger under a negative DC current bias,  the superconducting phase acts like a reverse diode in the DC current range $-I_c^- < I_{dc} < -I_c^+$ (highlighted by the shaded area in Fig.~\ref{figSC}e). Notably, the zero-field superconducting diode effect is non-volatile. As the DC current bias varies between $\pm 22$ nA, the sample alternatively switches between superconducting and normal state behavior  over the time span of more than an hour (Fig.~\ref{figSC}f).

The presence of either time-reversal, $\Theta$, or two-fold rotation symmetry, $C_{2z}$, implies $\Delta I_c=0$ (see SI 1 \cite{SI} for a symmetry analysis). As such, our observation of nonreciprocal behavior provides unambiguous evidence that both $C_{2z}$ and $\Theta$ are broken in the superconducting state. 
To provide additional evidence for the spontaneous breaking of $\Theta$ and its coexistence with superconductivity, we next demonstrate that the diode effect can be reversed through training with an external magnetic field.  Fig.~\ref{figdiode}a-b show that  the sign of $\Delta I_c$ correlates with the $B$-field direction in the presence of an out-of-plane magnetic field. As the magnetic field is slowly reduced to zero, the superconducting phase maintains the same nonreciprocity (Fig.~\ref{figdiode}c), giving rise to a hysteretic behavior where the sign of $\Delta I_c$ is determined by the field training history (Fig.~\ref{figdiode}d). After field training with positive (negative) $B$, the superconducting phase behaves like a reverse (forward) diode in the current range of $-I_c^- < I_{dc} < -I_c^+$ ($I_c^- < I_{dc} < I_c^+$). 
The fact that the zero-field nonreciprocity can be reversed through field training reveals that the underlying time-reversal-symmetry-breaking order can couple efficiently to an out-of-plane magnetic field, which provides important constraints on order parameter \cite{SI,DiodeTheoryPaper}: for instance, an in-plane spin polarization cannot be trained efficiently with an out-of-plane field, therefore, the underlying time-reversal symmetry breaking more likely arises from valley polarization or out-of-plane spin-polarization. 
Irrespective of the microscopic origin of the diode effect, it is an emergent property of the moir\'e system, since none of the component materials are magnetic.

Fig.~\ref{figdiode}e reveals an extra experimental knob to control the nonreciprocity of superconducting transport behavior without having to train the sample using an external magnetic field. After field training with a positive external magnetic field,  we show that the sign of the zero-field superconducting diode effect can be reliably and repeatedly switched by changing the charge carrier polarity via field-effect doping. 
This observation suggests that the spontaneous time-reversal symmetry breaking field underlying the diode effect remains robust as the sample is tuned outside of the superconducting density range. This  provides strong indication that time-reversal symmetry breaking arises from the underlying fermi surface of the normal phase ~\cite{DiodeTheoryPaper}. 
In addition, Fig.~\ref{figdiode}f shows that the superconducting diode effect persists throughout the entire density range of the superconducting phase. 
After field-training with a positive external magnetic field, the entire density range of the electron-doped (hole-doped) superconducting phase behaves like a reverse (forward) diode, and the sample exhibits dissipationless behavior  at $I_{dc}=-20$~nA ($I_{dc}=+52$~nA)  but appears highly resistive at $I_{dc}=+20$ nA ($I_{dc}=-52$~nA). 
The uniform response across the entire density range is consistent with a diode effect intrinsic to the superconducting phase, which is distinct from the nonreciprocity arising from a magnetic tunnel junction ~\cite{Diez2021magnetic} or Josephson networks ~\cite{Hooper2004}.
As such, we have realized a junction-free superconductor film, which operates as a supercurrent diode and achieves one-way transport of electrical charge without power consumption.
Notably, the density dependence of $\Delta I_c$, as shown in Fig.~\ref{figdiode}e, gives rise to
extreme nonreciprocity near the high density end of the superconducting regime. In this regime, the sample is resistive at $I_{dc}=0$ but exhibits dissipationless transport behavior when $I_{dc}$ is non-zero (Fig.~\ref{figdiode}g). Such extreme nonreciprocity provides further confirmation for a robust  diode effect that is intrinsic to the superconducting phase.

While the superconducting diode effect can be stabilized by training with a non-zero magnetic field, it can also be suppressed using a large DC current on the order of $\sim 200$ nA, as shown in Fig.~\ref{figT}a-b. The influence of the DC current likely originates from the dynamics of domain formation under a large current bias (see Fig.~\ref{fig:Current} and Fig.~\ref{fig:Currentmodel} for more discussions). 
Our ability to ``train'' and ``untrain'' the nonreciprocity offers a unique opportunity to examine transport behaviors that are directly associated with the superconducting diode effect.
For instance, the ``trained'' configuration exhibits an apparent increase in both longitudinal and Hall resistance around the transition temperature $T_c$, which is defined as the temperature where \Rxx\ in the untrained configuration diminishes. As superconductivity is destroyed at high temperature $T > T_c$, the influence of training and untraining diminishes  (Fig.~\ref{figT}e). 

This shows that the underlying time-reversal-symmetry-breaking order is only a rather weak perturbation to the band structure and, e.g., does not induce insulating behavior or completely polarize a flavor degree of three of the system (such as the valley or spin quantum number) at a temperature significantly larger than $T_c$; this is consistent with the fact that superconductivity can coexist with it, as required for the zero-field diode effect.
At the same time, the increase in \Rxy\ leads to an enhanced ratio of \Rxy$/$\Rxx\ compared to the normal state at high temperature (Fig.~\ref{figT}f). The enhancement in the ratio of \Rxy$/$\Rxx\ indicates that time-reversal, $\Theta$, or three-fold rotational symmetry, $C_{3z}$, is broken (see SI 4) \cite{SI}. Since the enhanced \Rxy/\Rxx\ disappears as the nonreciprocity is suppressed in the ``untrained'' configuration, the reduced symmetry is likely directly associated with the superconducting diode effect.

Notably, the superconducting behavior at small current bias is observed over a larger (smaller) density range  in the absence (presence) of the nonreciprocal behavior (Fig.~\ref{figT}c-d). 
It is worth pointing out that the variation in the superconducting density range results from the extreme nonreciprocity in the high density regime, where dissipationless transport is only robust with non-zero DC bias current (Fig.~\ref{figdiode}e and Fig.~\ref{figT}a-d). On the other hand, Fig.~\ref{fig:Current}b shows that $(I_++I_-)/2$ remains the same despite the onset of the zero-field diode effect, providing strong indication that the robustness of the superconducting phase is not influenced by the process of field training.  

Given the abundance of orbital ferromagnetism in graphene moir\'e systems, a natural explanation for the zero-field diode effect and its associated symmetry breaking is an interaction-induced imbalance in the valley occupation of the underlying fermi surface. The spontaneous formation of an imbalance of electrons in the two valleys breaks both $C_{2z}$ and $\Theta$, providing the necessary symmetry requirement for the emergence of a zero-field superconducting diode effect. As such,  a valley imbalanced fermi surface, in theory, can give rise to nonreciprocal superconducting transport behavior in the absence of proximity-induced SOC. However, the presence of (at least Ising) SOC is required for the valley order to couple to an external out-of-plane magnetic field (see Table~\ref{DiodeEffect}) \cite{SI,DiodeTheoryPaper}. In sample A, SOC could result from the proximity effect with the \WSe\ crystal, which explains the observed field trainability discussed above (see SI 5 and Fig.~\ref{fig:optDope} for more discussion regarding the influence of the proximity effect). 

An imbalance in the valley occupation of the underlying fermi surface has crucial consequences for the superconducting state \cite{DiodeTheoryPaper}. Let us denote the degree of valley imbalance by $V_z$, which can be qualitatively thought of as the difference in chemical potentials in the two valleys. 
At $V_z=0$, a valley balanced Fermi surface preserves time-reversal symmetry (top blue square in Fig.~\ref{figT}g). In this scenario, the two valleys  are related by reversing the sign of the crystalline momentum, $\vec{k}\rightarrow -\vec{k}$, making inter-valley pairing at zero center of mass momentum, $\vec{q}=0$, most favorable. In the presence of non-zero valley polarization, $V_z \neq 0$, this Kramers degeneracy is lifted. In general, a valley imbalanced fermi surface will reduce the  superconducting transition temperature. 
But if a superconducting phase survives, it is expected to exhibit zero-field diode effect.
Notably, there are two critical values in the valley polarization $V_z$. For a finite range of $V_z$ around zero, the associated Cooper pairs exhibit vanishing center of mass momentum $\vec{q}=0$ as a consequence of $C_{3z}$. This results in a critical value of valley imbalance, $V_{z1}$, below which $C_{3z}$ symmetry is preserved (blue region in Fig.~\ref{figT}g) and Cooper pairs form with zero center of mass momentum $\vec{q}=0$. At $V>V_{z1}$, valley imbalance gives rise to Cooper pairing with non-zero center of mass momentum appears, $\vec{q}\neq 0$. This finite-momentum pairing state also breaks $C_{3z}$. Given the extreme nonreciprocity in Fig.~\ref{figdiode}g, sample A likely falls inside this regime of valley imbalance, at least for certain $\nu_{tTLG}$. It is worth pointing out that the resulting nematicity also provides a natural explanation for the enhancement in \Rxy/\Rxx\ in the trained configuration (see SI 4) ~\cite{SI}. Given the partial valley imbalance, the observed variation in $\Delta I_c$ offers a direct characterization for the density dependence of $V_z$ in the underlying fermi surface. According to Fig.~\ref{figdiode}e and g, $V_z$ is maximized at large carrier density.

\begin{figure*}
\includegraphics[width=0.7\linewidth]{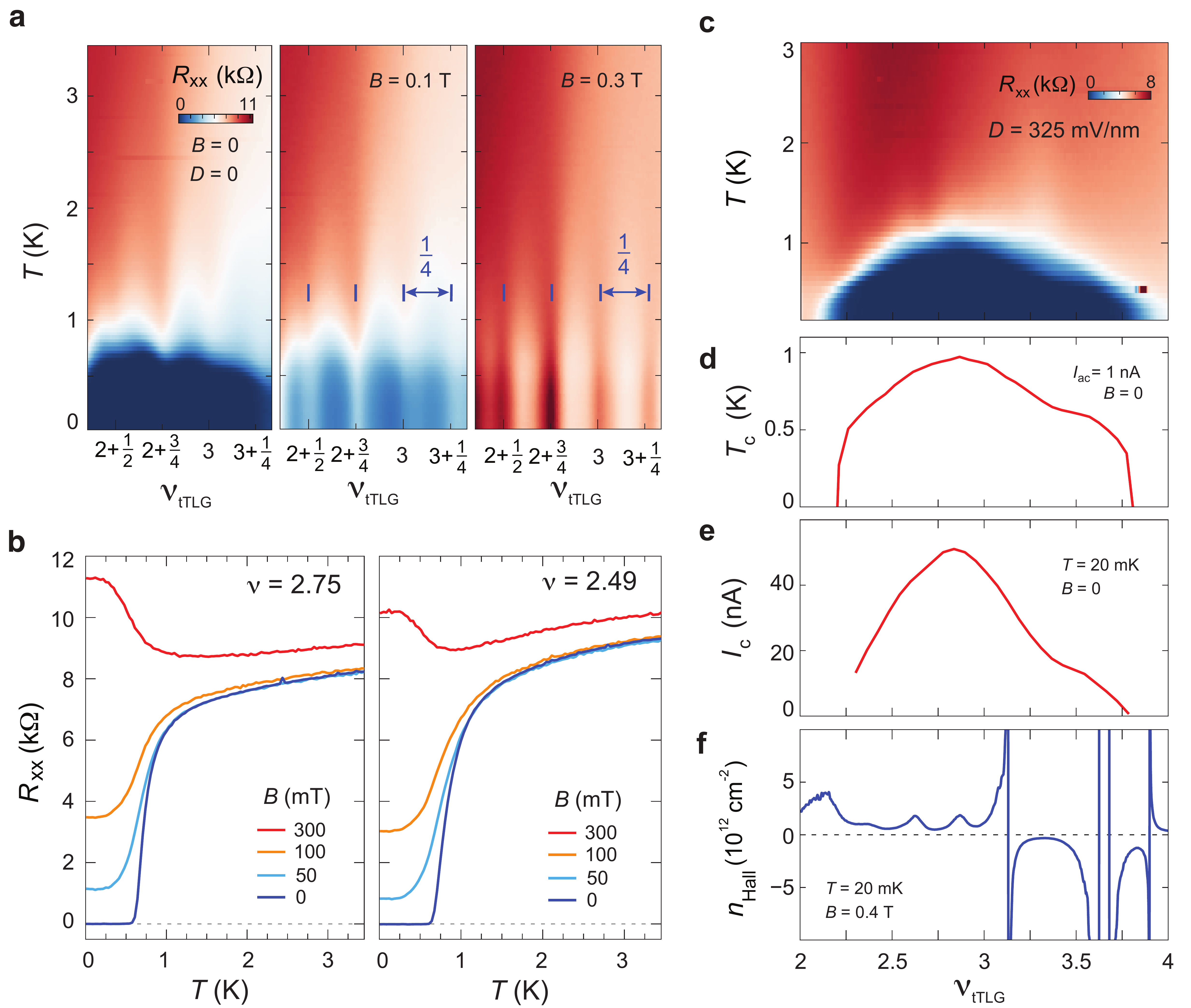}
\caption{\label{figDW} {\bf{Density waves and superconductivity.}}   (a) $\nu_{tTLG}-T$ map of \Rxx\ measured at different $B$. Vertical blue bars in the right panels mark $1/4$ period in moir\'e filling. (b)  \Rxx\ as a function of $T$ measured at $\nu= 2.75$ (left panel), $\nu=2.49$ (right panel) and different $B$. (c-f) are measured for the electron pocket at $D = 325$~mV/nm. (c) \Rxx \,as a function of temperature and filling. (d) Critical temperature extracted from (b). (e) Critical current at 20 mK and 0T. (f) Hall density at 20 mK, 0.4T.}
\end{figure*} 

Further increasing valley imbalance results in a fully valley polarized fermi surface at $V_z = V_{z2}$ (red square in Fig.~\ref{figT}g). A sufficiently large valley imbalance also requires Cooper pairs to form within the same valley. Such intra-valley pairing gives rise to extremely large center of mass momentum, of the scale of the Brillouin zone of the individual graphene layers, rendering the superconducting phase highly unstable. This offers a natural explanation for the absence of the zero-field diode effect near the magic angle (see Fig.~\ref{fig:Cdiode}). Owing to the dominating influence of flavor polarization near the magic angle, the fermi surface near $\nu=\pm2$ is either perfectly valley balanced $V_z=0$ or fully valley polarized $V_z = V_{z2}$ (red region in Fig.~\ref{figT}g). 
As a result, the observed superconducting phase near the magic angle is always associated with an underlying fermi surface which preserves time-reversal symmetry, and $\Delta I_c=0$ by symmetry.
Combined, the absence of the zero-field diode effect near the magic angle suggests that the suppressed influence of flavor polarization and partial imbalance in the valley occupation are key to stabilizing nonreciprocal supercurrents in the small twist angle regime ~\cite{DWPaper}.

Not withstanding the good agreement between the valley-polarization-based mechanism for the diode effect and the observed phenomenology, we note that, from a pure symmetry perspective, it is also possible that the normal state of the superconductor is non-magnetic but $\Theta$ is broken spontaneously at the superconducting transition \cite{DiodeTheoryPaper,2021arXiv211105340Z}. Indeed, there is exactly one candidate pairing state \cite{DiodeTheoryPaper} that our system can reach by a single continuous phase transition which has the correct symmetries. Transforming under a non-trivial representation of the point group ($E$ of $C_{3z}$), this state is expected to be fragile against impurities in the system and, hence, represents a less natural option than valley polarization. As it breaks $\Theta$, it can only be stabilized by fluctuations of a time-reversal-odd collective mode, such as spin fluctuations, and not by phonons alone \cite{PhysRevB.93.174509}; as such, future Coulomb screening experiments \cite{Liu2021DtBLG,Liu2021DtTLG} should be able to clarify the mechanism underlying the diode effect. 

In the following, we will investigate the relationship between superconductivity and density wave (DW) instabilities. Moir\'e energy bands are populated by density wave phases in the small twist angle regime~\cite{DWPaper} in a range of $\nu_{tTLG}$ and $D$ that overlaps with that of superconductivity in Fig.~\ref{figSC}c. 
This is demonstrated in Fig.~\ref{figDW}a by the density modulation of 1/4 moir\'e filling in \Rxx\ as superconductivity is suppressed by an external magnetic field $B$.
Interestingly, the melting temperature of the DW coincides approximately with $T_c$ of the superconducting phase (Fig.~\ref{figDW}b), suggesting that the underlying energy scale is comparable between superconductivity and DW order. 
As such, it is important to address the connection between these two phenomena.
In particular, we have to distinguish between two different scenarios: (i) DW order and superconductivity coexist microscopically, i.e., upon decreasing $B$, Cooper pairs form in the mini bands reconstructed by the DW order; (ii) superconductivity and DW order are competing ground states and DW order disappears at the onset of superconductivity (and vice versa). 
To this end, we will compare the density dependence of superconductivity at $B=0$ and of the DW states at $B=0.4$ T. At $D=325$ mV/nm, the Hall density measured at $B=0.4$ T exhibits a complex density modulation of $1/4$ moir\'e filling, with several divergences in the range $3<\nu_{tTLG}<4$ (Fig.~\ref{figDW}f), which are associated with the DW-induced band reconstruction~\cite{DWPaper}. 
Regardless of its pairing symmetry, the strength of superconductivity, reflected by quantities such as $T_c$ and $I_c$, is generically expected to be sensitive to changes in the density of states (DOS) of the parent Fermi surface. 
However, the critical temperature $T_c$ and critical current $I_c$ of the superconducting phase at $B=0$ vary smoothly with $\nu_{tTLG}$ (Fig.~\ref{figDW}d-e). The fact that the superconducting phase is largely unaffected by DW-induced changes in the DOS indicates that DW states at $0.4$ T are not the parent state of superconductivity. This favors scenario (ii) over (i) and superconductivity and DW are more likely competing orders. 
In this case, the magnetic order responsible for the diode effect, such as the valley polarization discussed above, is an additional instability, distinct from DW order. 
We note that superconductivity and DW phases may coexist at the base temperature of $T \sim 20$ mK in certain parts of the phase space, which gives rise to extreme anisotropic transport behavior in the superconducting phase (see Supplemental Fig.~\ref{f:nematic2}). 

Lastly, we will comment on the twist angle dependence of the superconducting phase in tTLG \cite{Park2021tTLG,Hao2021tTLG,Liu2021DtTLG,DWPaper}. 
As shown in a recent experimental work \cite{DWPaper}, the interplay between Coulomb correlation and superconductivity in the small twist angle regime takes a form distinct from the magic angle. In the small twist angle regime, which is marked by the red shaded area in Fig.~\ref{figSC}a, a lack of hierarchy between correlation-driven phases 
indicates that the influence of flavor polarization is substantially diminished ~\cite{DWPaper}. This provides a key ingredient for realizing the zero-field diode effect: a partial imbalance in valley occupation of the underlying fermi surface. On the other hand, partial valley polarization is found to be unfavorable near the magic angle, owing to the dominating influence of flavor polarization. Combined, this provides a natural explanation for twist angle dependence of the zero-field diode effect we find. Taken together, this establishes the small-twist-angle regime in twisted trilayer graphene as a novel paradigm for correlated physics, which involves the complex interplay of superconductivity, broken time-reversal symmetry, finite momentum pairing, and density-wave instabilities.

\section*{Acknowledgments}
This work was primarily supported by Brown University. Device fabrication was performed in the Institute for Molecular and Nanoscale Innovation at Brown University. The authors acknowledge the use of equipment funded by the MRI award DMR-1827453. P.S. acknowledges support from the Brown University Undergraduate Teaching and Research Awards.
H. D. Scammell acknowledges funding from ARC Centre of Excellence FLEET.
K.W. and T.T. acknowledge support from the Elemental Strategy Initiative
conducted by the MEXT, Japan (Grant Number JPMXP0112101001) and  JSPS
KAKENHI (Grant Numbers 19H05790, 20H00354 and 21H05233).

\bibliography{Li_ref}

\begin{thebibliography}{43}%
\makeatletter
\providecommand \@ifxundefined [1]{%
 \@ifx{#1\undefined}
}%
\providecommand \@ifnum [1]{%
 \ifnum #1\expandafter \@firstoftwo
 \else \expandafter \@secondoftwo
 \fi
}%
\providecommand \@ifx [1]{%
 \ifx #1\expandafter \@firstoftwo
 \else \expandafter \@secondoftwo
 \fi
}%
\providecommand \natexlab [1]{#1}%
\providecommand \enquote  [1]{``#1''}%
\providecommand \bibnamefont  [1]{#1}%
\providecommand \bibfnamefont [1]{#1}%
\providecommand \citenamefont [1]{#1}%
\providecommand \href@noop [0]{\@secondoftwo}%
\providecommand \href [0]{\begingroup \@sanitize@url \@href}%
\providecommand \@href[1]{\@@startlink{#1}\@@href}%
\providecommand \@@href[1]{\endgroup#1\@@endlink}%
\providecommand \@sanitize@url [0]{\catcode `\\12\catcode `\$12\catcode
  `\&12\catcode `\#12\catcode `\^12\catcode `\_12\catcode `\%12\relax}%
\providecommand \@@startlink[1]{}%
\providecommand \@@endlink[0]{}%
\providecommand \url  [0]{\begingroup\@sanitize@url \@url }%
\providecommand \@url [1]{\endgroup\@href {#1}{\urlprefix }}%
\providecommand \urlprefix  [0]{URL }%
\providecommand \Eprint [0]{\href }%
\providecommand \doibase [0]{http://dx.doi.org/}%
\providecommand \selectlanguage [0]{\@gobble}%
\providecommand \bibinfo  [0]{\@secondoftwo}%
\providecommand \bibfield  [0]{\@secondoftwo}%
\providecommand \translation [1]{[#1]}%
\providecommand \BibitemOpen [0]{}%
\providecommand \bibitemStop [0]{}%
\providecommand \bibitemNoStop [0]{.\EOS\space}%
\providecommand \EOS [0]{\spacefactor3000\relax}%
\providecommand \BibitemShut  [1]{\csname bibitem#1\endcsname}%
\let\auto@bib@innerbib\@empty
\bibitem [{\citenamefont {Ando}\ \emph {et~al.}(2020)\citenamefont {Ando},
  \citenamefont {Miyasaka}, \citenamefont {Li}, \citenamefont {Ishizuka},
  \citenamefont {Arakawa}, \citenamefont {Shiota}, \citenamefont {Moriyama},
  \citenamefont {Yanase},\ and\ \citenamefont {Ono}}]{Ando2020diodes}%
  \BibitemOpen
  \bibfield  {author} {\bibinfo {author} {\bibfnamefont {F.}~\bibnamefont
  {Ando}}, \bibinfo {author} {\bibfnamefont {Y.}~\bibnamefont {Miyasaka}},
  \bibinfo {author} {\bibfnamefont {T.}~\bibnamefont {Li}}, \bibinfo {author}
  {\bibfnamefont {J.}~\bibnamefont {Ishizuka}}, \bibinfo {author}
  {\bibfnamefont {T.}~\bibnamefont {Arakawa}}, \bibinfo {author} {\bibfnamefont
  {Y.}~\bibnamefont {Shiota}}, \bibinfo {author} {\bibfnamefont
  {T.}~\bibnamefont {Moriyama}}, \bibinfo {author} {\bibfnamefont
  {Y.}~\bibnamefont {Yanase}}, \ and\ \bibinfo {author} {\bibfnamefont
  {T.}~\bibnamefont {Ono}},\ }\href@noop {} {\bibfield  {journal} {\bibinfo
  {journal} {Nature}\ }\textbf {\bibinfo {volume} {584}},\ \bibinfo {pages}
  {373} (\bibinfo {year} {2020})}\BibitemShut {NoStop}%
\bibitem [{\citenamefont {Diez-Merida}\ \emph {et~al.}(2021)\citenamefont
  {Diez-Merida}, \citenamefont {Diez-Carlon}, \citenamefont {Yang},
  \citenamefont {Xie}, \citenamefont {Gao}, \citenamefont {Watanabe},
  \citenamefont {Taniguchi}, \citenamefont {Lu}, \citenamefont {Law},\ and\
  \citenamefont {Efetov}}]{Diez2021magnetic}%
  \BibitemOpen
  \bibfield  {author} {\bibinfo {author} {\bibfnamefont {J.}~\bibnamefont
  {Diez-Merida}}, \bibinfo {author} {\bibfnamefont {A.}~\bibnamefont
  {Diez-Carlon}}, \bibinfo {author} {\bibfnamefont {S.}~\bibnamefont {Yang}},
  \bibinfo {author} {\bibfnamefont {Y.-M.}\ \bibnamefont {Xie}}, \bibinfo
  {author} {\bibfnamefont {X.-J.}\ \bibnamefont {Gao}}, \bibinfo {author}
  {\bibfnamefont {K.}~\bibnamefont {Watanabe}}, \bibinfo {author}
  {\bibfnamefont {T.}~\bibnamefont {Taniguchi}}, \bibinfo {author}
  {\bibfnamefont {X.}~\bibnamefont {Lu}}, \bibinfo {author} {\bibfnamefont
  {K.}~\bibnamefont {Law}}, \ and\ \bibinfo {author} {\bibfnamefont {D.~K.}\
  \bibnamefont {Efetov}},\ }\href@noop {} {\bibfield  {journal} {\bibinfo
  {journal} {arXiv preprint arXiv:2110.01067}\ } (\bibinfo {year}
  {2021})}\BibitemShut {NoStop}%
\bibitem [{\citenamefont {Park}\ \emph {et~al.}(2021)\citenamefont {Park},
  \citenamefont {Cao}, \citenamefont {Watanabe}, \citenamefont {Taniguchi},\
  and\ \citenamefont {Jarillo-Herrero}}]{Park2021tTLG}%
  \BibitemOpen
  \bibfield  {author} {\bibinfo {author} {\bibfnamefont {J.~M.}\ \bibnamefont
  {Park}}, \bibinfo {author} {\bibfnamefont {Y.}~\bibnamefont {Cao}}, \bibinfo
  {author} {\bibfnamefont {K.}~\bibnamefont {Watanabe}}, \bibinfo {author}
  {\bibfnamefont {T.}~\bibnamefont {Taniguchi}}, \ and\ \bibinfo {author}
  {\bibfnamefont {P.}~\bibnamefont {Jarillo-Herrero}},\ }\href@noop {}
  {\bibfield  {journal} {\bibinfo  {journal} {Nature}\ }\textbf {\bibinfo
  {volume} {590}},\ \bibinfo {pages} {249} (\bibinfo {year}
  {2021})}\BibitemShut {NoStop}%
\bibitem [{\citenamefont {Ginzburg}(1957)}]{Ginzburg1957ferromagnetic}%
  \BibitemOpen
  \bibfield  {author} {\bibinfo {author} {\bibfnamefont {V.}~\bibnamefont
  {Ginzburg}},\ }\href@noop {} {\bibfield  {journal} {\bibinfo  {journal}
  {Soviet Physics Jetp-Ussr}\ }\textbf {\bibinfo {volume} {4}},\ \bibinfo
  {pages} {153} (\bibinfo {year} {1957})}\BibitemShut {NoStop}%
\bibitem [{\citenamefont {Matthias}\ \emph {et~al.}(1958)\citenamefont
  {Matthias}, \citenamefont {Suhl},\ and\ \citenamefont
  {Corenzwit}}]{Matthias1958spin}%
  \BibitemOpen
  \bibfield  {author} {\bibinfo {author} {\bibfnamefont {B.}~\bibnamefont
  {Matthias}}, \bibinfo {author} {\bibfnamefont {H.}~\bibnamefont {Suhl}}, \
  and\ \bibinfo {author} {\bibfnamefont {E.}~\bibnamefont {Corenzwit}},\
  }\href@noop {} {\bibfield  {journal} {\bibinfo  {journal} {Physical Review
  Letters}\ }\textbf {\bibinfo {volume} {1}},\ \bibinfo {pages} {92} (\bibinfo
  {year} {1958})}\BibitemShut {NoStop}%
\bibitem [{\citenamefont {Anderson}(1959)}]{ANDERSON195926}%
  \BibitemOpen
  \bibfield  {author} {\bibinfo {author} {\bibfnamefont {P.}~\bibnamefont
  {Anderson}},\ }\href {\doibase https://doi.org/10.1016/0022-3697(59)90036-8}
  {\bibfield  {journal} {\bibinfo  {journal} {Journal of Physics and Chemistry
  of Solids}\ }\textbf {\bibinfo {volume} {11}},\ \bibinfo {pages} {26}
  (\bibinfo {year} {1959})}\BibitemShut {NoStop}%
\bibitem [{\citenamefont {Fulde}\ and\ \citenamefont
  {Ferrell}(1964)}]{Fulde1964FFLO}%
  \BibitemOpen
  \bibfield  {author} {\bibinfo {author} {\bibfnamefont {P.}~\bibnamefont
  {Fulde}}\ and\ \bibinfo {author} {\bibfnamefont {R.~A.}\ \bibnamefont
  {Ferrell}},\ }\href@noop {} {\bibfield  {journal} {\bibinfo  {journal}
  {Physical Review}\ }\textbf {\bibinfo {volume} {135}},\ \bibinfo {pages}
  {A550} (\bibinfo {year} {1964})}\BibitemShut {NoStop}%
\bibitem [{\citenamefont {Larkin}\ and\ \citenamefont
  {Ovchinnikov}(1965)}]{Larkin1965FFLO}%
  \BibitemOpen
  \bibfield  {author} {\bibinfo {author} {\bibfnamefont {A.}~\bibnamefont
  {Larkin}}\ and\ \bibinfo {author} {\bibfnamefont {Y.~N.}\ \bibnamefont
  {Ovchinnikov}},\ }\href@noop {} {\bibfield  {journal} {\bibinfo  {journal}
  {Soviet Physics-JETP}\ }\textbf {\bibinfo {volume} {20}},\ \bibinfo {pages}
  {762} (\bibinfo {year} {1965})}\BibitemShut {NoStop}%
\bibitem [{\citenamefont {Buzdin}(2005)}]{Buzdin2005proximity}%
  \BibitemOpen
  \bibfield  {author} {\bibinfo {author} {\bibfnamefont {A.~I.}\ \bibnamefont
  {Buzdin}},\ }\href@noop {} {\bibfield  {journal} {\bibinfo  {journal}
  {Reviews of modern physics}\ }\textbf {\bibinfo {volume} {77}},\ \bibinfo
  {pages} {935} (\bibinfo {year} {2005})}\BibitemShut {NoStop}%
\bibitem [{\citenamefont {Bianchi}\ \emph {et~al.}(2003)\citenamefont
  {Bianchi}, \citenamefont {Movshovich}, \citenamefont {Capan}, \citenamefont
  {Pagliuso},\ and\ \citenamefont {Sarrao}}]{PhysRevLett.91.187004}%
  \BibitemOpen
  \bibfield  {author} {\bibinfo {author} {\bibfnamefont {A.}~\bibnamefont
  {Bianchi}}, \bibinfo {author} {\bibfnamefont {R.}~\bibnamefont {Movshovich}},
  \bibinfo {author} {\bibfnamefont {C.}~\bibnamefont {Capan}}, \bibinfo
  {author} {\bibfnamefont {P.~G.}\ \bibnamefont {Pagliuso}}, \ and\ \bibinfo
  {author} {\bibfnamefont {J.~L.}\ \bibnamefont {Sarrao}},\ }\href {\doibase
  10.1103/PhysRevLett.91.187004} {\bibfield  {journal} {\bibinfo  {journal}
  {Phys. Rev. Lett.}\ }\textbf {\bibinfo {volume} {91}},\ \bibinfo {pages}
  {187004} (\bibinfo {year} {2003})}\BibitemShut {NoStop}%
\bibitem [{\citenamefont {Radovan}\ \emph {et~al.}(2003)\citenamefont
  {Radovan}, \citenamefont {Fortune}, \citenamefont {Murphy}, \citenamefont
  {Hannahs}, \citenamefont {Palm}, \citenamefont {Tozer},\ and\ \citenamefont
  {Hall}}]{FFLO2}%
  \BibitemOpen
  \bibfield  {author} {\bibinfo {author} {\bibfnamefont {H.~A.}\ \bibnamefont
  {Radovan}}, \bibinfo {author} {\bibfnamefont {N.~A.}\ \bibnamefont
  {Fortune}}, \bibinfo {author} {\bibfnamefont {T.~P.}\ \bibnamefont {Murphy}},
  \bibinfo {author} {\bibfnamefont {S.~T.}\ \bibnamefont {Hannahs}}, \bibinfo
  {author} {\bibfnamefont {E.~C.}\ \bibnamefont {Palm}}, \bibinfo {author}
  {\bibfnamefont {S.~W.}\ \bibnamefont {Tozer}}, \ and\ \bibinfo {author}
  {\bibfnamefont {D.}~\bibnamefont {Hall}},\ }\href {\doibase
  10.1038/nature01842} {\bibfield  {journal} {\bibinfo  {journal} {Nature}\
  }\textbf {\bibinfo {volume} {425}},\ \bibinfo {pages} {51} (\bibinfo {year}
  {2003})}\BibitemShut {NoStop}%
\bibitem [{\citenamefont {Felner}\ \emph {et~al.}(1997)\citenamefont {Felner},
  \citenamefont {Asaf}, \citenamefont {Levi},\ and\ \citenamefont
  {Millo}}]{Coexistence1}%
  \BibitemOpen
  \bibfield  {author} {\bibinfo {author} {\bibfnamefont {I.}~\bibnamefont
  {Felner}}, \bibinfo {author} {\bibfnamefont {U.}~\bibnamefont {Asaf}},
  \bibinfo {author} {\bibfnamefont {Y.}~\bibnamefont {Levi}}, \ and\ \bibinfo
  {author} {\bibfnamefont {O.}~\bibnamefont {Millo}},\ }\href {\doibase
  10.1103/PhysRevB.55.R3374} {\bibfield  {journal} {\bibinfo  {journal} {Phys.
  Rev. B}\ }\textbf {\bibinfo {volume} {55}},\ \bibinfo {pages} {R3374}
  (\bibinfo {year} {1997})}\BibitemShut {NoStop}%
\bibitem [{\citenamefont {Aoki}\ \emph {et~al.}(2001)\citenamefont {Aoki},
  \citenamefont {Huxley}, \citenamefont {Ressouche}, \citenamefont
  {Braithwaite}, \citenamefont {Flouquet}, \citenamefont {Brison},
  \citenamefont {Lhotel},\ and\ \citenamefont {Paulsen}}]{Coexistence2}%
  \BibitemOpen
  \bibfield  {author} {\bibinfo {author} {\bibfnamefont {D.}~\bibnamefont
  {Aoki}}, \bibinfo {author} {\bibfnamefont {A.}~\bibnamefont {Huxley}},
  \bibinfo {author} {\bibfnamefont {E.}~\bibnamefont {Ressouche}}, \bibinfo
  {author} {\bibfnamefont {D.}~\bibnamefont {Braithwaite}}, \bibinfo {author}
  {\bibfnamefont {J.}~\bibnamefont {Flouquet}}, \bibinfo {author}
  {\bibfnamefont {J.-P.}\ \bibnamefont {Brison}}, \bibinfo {author}
  {\bibfnamefont {E.}~\bibnamefont {Lhotel}}, \ and\ \bibinfo {author}
  {\bibfnamefont {C.}~\bibnamefont {Paulsen}},\ }\href {\doibase
  10.1038/35098048} {\bibfield  {journal} {\bibinfo  {journal} {Nature}\
  }\textbf {\bibinfo {volume} {413}},\ \bibinfo {pages} {613} (\bibinfo {year}
  {2001})}\BibitemShut {NoStop}%
\bibitem [{\citenamefont {Pfleiderer}\ \emph {et~al.}(2001)\citenamefont
  {Pfleiderer}, \citenamefont {Uhlarz}, \citenamefont {Hayden}, \citenamefont
  {Vollmer}, \citenamefont {L{\"o}hneysen}, \citenamefont {Bernhoeft},\ and\
  \citenamefont {Lonzarich}}]{Coexistence3}%
  \BibitemOpen
  \bibfield  {author} {\bibinfo {author} {\bibfnamefont {C.}~\bibnamefont
  {Pfleiderer}}, \bibinfo {author} {\bibfnamefont {M.}~\bibnamefont {Uhlarz}},
  \bibinfo {author} {\bibfnamefont {S.~M.}\ \bibnamefont {Hayden}}, \bibinfo
  {author} {\bibfnamefont {R.}~\bibnamefont {Vollmer}}, \bibinfo {author}
  {\bibfnamefont {H.~v.}\ \bibnamefont {L{\"o}hneysen}}, \bibinfo {author}
  {\bibfnamefont {N.~R.}\ \bibnamefont {Bernhoeft}}, \ and\ \bibinfo {author}
  {\bibfnamefont {G.~G.}\ \bibnamefont {Lonzarich}},\ }\href {\doibase
  10.1038/35083531} {\bibfield  {journal} {\bibinfo  {journal} {Nature}\
  }\textbf {\bibinfo {volume} {412}},\ \bibinfo {pages} {58} (\bibinfo {year}
  {2001})}\BibitemShut {NoStop}%
\bibitem [{\citenamefont {Ren}\ \emph {et~al.}(2009)\citenamefont {Ren},
  \citenamefont {Tao}, \citenamefont {Jiang}, \citenamefont {Feng},
  \citenamefont {Wang}, \citenamefont {Dai}, \citenamefont {Cao},\ and\
  \citenamefont {Xu}}]{Ren2009coexistence}%
  \BibitemOpen
  \bibfield  {author} {\bibinfo {author} {\bibfnamefont {Z.}~\bibnamefont
  {Ren}}, \bibinfo {author} {\bibfnamefont {Q.}~\bibnamefont {Tao}}, \bibinfo
  {author} {\bibfnamefont {S.}~\bibnamefont {Jiang}}, \bibinfo {author}
  {\bibfnamefont {C.}~\bibnamefont {Feng}}, \bibinfo {author} {\bibfnamefont
  {C.}~\bibnamefont {Wang}}, \bibinfo {author} {\bibfnamefont {J.}~\bibnamefont
  {Dai}}, \bibinfo {author} {\bibfnamefont {G.}~\bibnamefont {Cao}}, \ and\
  \bibinfo {author} {\bibfnamefont {Z.}~\bibnamefont {Xu}},\ }\href {\doibase
  10.1103/PhysRevLett.102.137002} {\bibfield  {journal} {\bibinfo  {journal}
  {Phys. Rev. Lett.}\ }\textbf {\bibinfo {volume} {102}},\ \bibinfo {pages}
  {137002} (\bibinfo {year} {2009})}\BibitemShut {NoStop}%
\bibitem [{\citenamefont {Dikin}\ \emph {et~al.}(2011)\citenamefont {Dikin},
  \citenamefont {Mehta}, \citenamefont {Bark}, \citenamefont {Folkman},
  \citenamefont {Eom},\ and\ \citenamefont
  {Chandrasekhar}}]{Dikin2011coexistence}%
  \BibitemOpen
  \bibfield  {author} {\bibinfo {author} {\bibfnamefont {D.~A.}\ \bibnamefont
  {Dikin}}, \bibinfo {author} {\bibfnamefont {M.}~\bibnamefont {Mehta}},
  \bibinfo {author} {\bibfnamefont {C.~W.}\ \bibnamefont {Bark}}, \bibinfo
  {author} {\bibfnamefont {C.~M.}\ \bibnamefont {Folkman}}, \bibinfo {author}
  {\bibfnamefont {C.~B.}\ \bibnamefont {Eom}}, \ and\ \bibinfo {author}
  {\bibfnamefont {V.}~\bibnamefont {Chandrasekhar}},\ }\href {\doibase
  10.1103/PhysRevLett.107.056802} {\bibfield  {journal} {\bibinfo  {journal}
  {Phys. Rev. Lett.}\ }\textbf {\bibinfo {volume} {107}},\ \bibinfo {pages}
  {056802} (\bibinfo {year} {2011})}\BibitemShut {NoStop}%
\bibitem [{\citenamefont {Li}\ \emph {et~al.}(2011)\citenamefont {Li},
  \citenamefont {Richter}, \citenamefont {Mannhart},\ and\ \citenamefont
  {Ashoori}}]{LuLi2011coexistence}%
  \BibitemOpen
  \bibfield  {author} {\bibinfo {author} {\bibfnamefont {L.}~\bibnamefont
  {Li}}, \bibinfo {author} {\bibfnamefont {C.}~\bibnamefont {Richter}},
  \bibinfo {author} {\bibfnamefont {J.}~\bibnamefont {Mannhart}}, \ and\
  \bibinfo {author} {\bibfnamefont {R.}~\bibnamefont {Ashoori}},\ }\href@noop
  {} {\bibfield  {journal} {\bibinfo  {journal} {Nature physics}\ }\textbf
  {\bibinfo {volume} {7}},\ \bibinfo {pages} {762} (\bibinfo {year}
  {2011})}\BibitemShut {NoStop}%
\bibitem [{\citenamefont {Bert}\ \emph {et~al.}(2011)\citenamefont {Bert},
  \citenamefont {Kalisky}, \citenamefont {Bell}, \citenamefont {Kim},
  \citenamefont {Hikita}, \citenamefont {Hwang},\ and\ \citenamefont
  {Moler}}]{Bert2011direct}%
  \BibitemOpen
  \bibfield  {author} {\bibinfo {author} {\bibfnamefont {J.~A.}\ \bibnamefont
  {Bert}}, \bibinfo {author} {\bibfnamefont {B.}~\bibnamefont {Kalisky}},
  \bibinfo {author} {\bibfnamefont {C.}~\bibnamefont {Bell}}, \bibinfo {author}
  {\bibfnamefont {M.}~\bibnamefont {Kim}}, \bibinfo {author} {\bibfnamefont
  {Y.}~\bibnamefont {Hikita}}, \bibinfo {author} {\bibfnamefont {H.~Y.}\
  \bibnamefont {Hwang}}, \ and\ \bibinfo {author} {\bibfnamefont {K.~A.}\
  \bibnamefont {Moler}},\ }\href@noop {} {\bibfield  {journal} {\bibinfo
  {journal} {Nature physics}\ }\textbf {\bibinfo {volume} {7}},\ \bibinfo
  {pages} {767} (\bibinfo {year} {2011})}\BibitemShut {NoStop}%
\bibitem [{\citenamefont {{Yuan}}\ and\ \citenamefont
  {{Fu}}(2021)}]{Yuan2021diodes}%
  \BibitemOpen
  \bibfield  {author} {\bibinfo {author} {\bibfnamefont {N.~F.~Q.}\
  \bibnamefont {{Yuan}}}\ and\ \bibinfo {author} {\bibfnamefont
  {L.}~\bibnamefont {{Fu}}},\ }\href@noop {} {\bibfield  {journal} {\bibinfo
  {journal} {arXiv e-prints}\ } (\bibinfo {year} {2021})},\ \Eprint
  {http://arxiv.org/abs/2106.01909} {arXiv:2106.01909 [cond-mat.supr-con]}
  \BibitemShut {NoStop}%
\bibitem [{\citenamefont {{Daido}}\ \emph {et~al.}(2021)\citenamefont
  {{Daido}}, \citenamefont {{Ikeda}},\ and\ \citenamefont
  {{Yanase}}}]{DaidoSCDiode}%
  \BibitemOpen
  \bibfield  {author} {\bibinfo {author} {\bibfnamefont {A.}~\bibnamefont
  {{Daido}}}, \bibinfo {author} {\bibfnamefont {Y.}~\bibnamefont {{Ikeda}}}, \
  and\ \bibinfo {author} {\bibfnamefont {Y.}~\bibnamefont {{Yanase}}},\
  }\href@noop {} {\bibfield  {journal} {\bibinfo  {journal} {arXiv e-prints}\ }
  (\bibinfo {year} {2021})},\ \Eprint {http://arxiv.org/abs/2106.03326}
  {arXiv:2106.03326 [cond-mat.supr-con]} \BibitemShut {NoStop}%
\bibitem [{\citenamefont {{He}}\ \emph {et~al.}(2021)\citenamefont {{He}},
  \citenamefont {{Tanaka}},\ and\ \citenamefont {{Nagaosa}}}]{HeSCDiode}%
  \BibitemOpen
  \bibfield  {author} {\bibinfo {author} {\bibfnamefont {J.~J.}\ \bibnamefont
  {{He}}}, \bibinfo {author} {\bibfnamefont {Y.}~\bibnamefont {{Tanaka}}}, \
  and\ \bibinfo {author} {\bibfnamefont {N.}~\bibnamefont {{Nagaosa}}},\
  }\href@noop {} {\bibfield  {journal} {\bibinfo  {journal} {arXiv e-prints}\ }
  (\bibinfo {year} {2021})},\ \Eprint {http://arxiv.org/abs/2106.03575}
  {arXiv:2106.03575 [cond-mat.supr-con]} \BibitemShut {NoStop}%
\bibitem [{\citenamefont {Ichikawa}\ \emph {et~al.}(1994)\citenamefont
  {Ichikawa}, \citenamefont {Nishizaki}, \citenamefont {Yamabe}, \citenamefont
  {Yamasaki}, \citenamefont {Fukami},\ and\ \citenamefont
  {Aomine}}]{Ichikawa1994IV}%
  \BibitemOpen
  \bibfield  {author} {\bibinfo {author} {\bibfnamefont {F.}~\bibnamefont
  {Ichikawa}}, \bibinfo {author} {\bibfnamefont {T.}~\bibnamefont {Nishizaki}},
  \bibinfo {author} {\bibfnamefont {K.}~\bibnamefont {Yamabe}}, \bibinfo
  {author} {\bibfnamefont {Y.}~\bibnamefont {Yamasaki}}, \bibinfo {author}
  {\bibfnamefont {T.}~\bibnamefont {Fukami}}, \ and\ \bibinfo {author}
  {\bibfnamefont {T.}~\bibnamefont {Aomine}},\ }\href@noop {} {\bibfield
  {journal} {\bibinfo  {journal} {Physica C: Superconductivity}\ }\textbf
  {\bibinfo {volume} {235}},\ \bibinfo {pages} {3095} (\bibinfo {year}
  {1994})}\BibitemShut {NoStop}%
\bibitem [{\citenamefont {Jiang}\ \emph {et~al.}(1994)\citenamefont {Jiang},
  \citenamefont {Connolly}, \citenamefont {Hagen},\ and\ \citenamefont
  {Lobb}}]{Jiang1994IV}%
  \BibitemOpen
  \bibfield  {author} {\bibinfo {author} {\bibfnamefont {X.}~\bibnamefont
  {Jiang}}, \bibinfo {author} {\bibfnamefont {P.}~\bibnamefont {Connolly}},
  \bibinfo {author} {\bibfnamefont {S.}~\bibnamefont {Hagen}}, \ and\ \bibinfo
  {author} {\bibfnamefont {C.}~\bibnamefont {Lobb}},\ }\href@noop {} {\bibfield
   {journal} {\bibinfo  {journal} {Physical Review B}\ }\textbf {\bibinfo
  {volume} {49}},\ \bibinfo {pages} {9244} (\bibinfo {year}
  {1994})}\BibitemShut {NoStop}%
\bibitem [{\citenamefont {Broussard}\ and\ \citenamefont
  {Geballe}(1988)}]{Broussard1988IV}%
  \BibitemOpen
  \bibfield  {author} {\bibinfo {author} {\bibfnamefont {P.}~\bibnamefont
  {Broussard}}\ and\ \bibinfo {author} {\bibfnamefont {T.}~\bibnamefont
  {Geballe}},\ }\href@noop {} {\bibfield  {journal} {\bibinfo  {journal}
  {Physical Review B}\ }\textbf {\bibinfo {volume} {37}},\ \bibinfo {pages}
  {68} (\bibinfo {year} {1988})}\BibitemShut {NoStop}%
\bibitem [{SI()}]{SI}%
  \BibitemOpen
  \href@noop {} {\bibinfo  {journal} {Please see the supplementary materials}\
  }\BibitemShut {NoStop}%
\bibitem [{\citenamefont {Scammell}\ \emph {et~al.}(2021)\citenamefont
  {Scammell}, \citenamefont {Li},\ and\ \citenamefont
  {Scheurer}}]{DiodeTheoryPaper}%
  \BibitemOpen
\bibfield  {journal} {  }\bibfield  {author} {\bibinfo {author} {\bibfnamefont
  {H.~D.}\ \bibnamefont {Scammell}}, \bibinfo {author} {\bibfnamefont
  {J.}~\bibnamefont {Li}}, \ and\ \bibinfo {author} {\bibfnamefont {M.~S.}\
  \bibnamefont {Scheurer}},\ }\href@noop {} {\bibfield  {journal} {\bibinfo
  {journal} {arXiv e-prints}\ } (\bibinfo {year} {2021})},\ \Eprint
  {http://arxiv.org/abs/2112.09115} {arXiv:2112.09115 [cond-mat.str-el]}
  \BibitemShut {NoStop}%
\bibitem [{\citenamefont {Hooper}\ \emph {et~al.}(2004)\citenamefont {Hooper},
  \citenamefont {Mao}, \citenamefont {Nelson}, \citenamefont {Liu},
  \citenamefont {Wada},\ and\ \citenamefont {Maeno}}]{Hooper2004}%
  \BibitemOpen
  \bibfield  {author} {\bibinfo {author} {\bibfnamefont {J.}~\bibnamefont
  {Hooper}}, \bibinfo {author} {\bibfnamefont {Z.~Q.}\ \bibnamefont {Mao}},
  \bibinfo {author} {\bibfnamefont {K.~D.}\ \bibnamefont {Nelson}}, \bibinfo
  {author} {\bibfnamefont {Y.}~\bibnamefont {Liu}}, \bibinfo {author}
  {\bibfnamefont {M.}~\bibnamefont {Wada}}, \ and\ \bibinfo {author}
  {\bibfnamefont {Y.}~\bibnamefont {Maeno}},\ }\href {\doibase
  10.1103/PhysRevB.70.014510} {\bibfield  {journal} {\bibinfo  {journal} {Phys.
  Rev. B}\ }\textbf {\bibinfo {volume} {70}},\ \bibinfo {pages} {014510}
  (\bibinfo {year} {2004})}\BibitemShut {NoStop}%
\bibitem [{\citenamefont {Siriviboon}\ \emph {et~al.}(2021)\citenamefont
  {Siriviboon}, \citenamefont {Lin}, \citenamefont {Scammell}, \citenamefont
  {Liu}, \citenamefont {Rhodes}, \citenamefont {Watanabe}, \citenamefont
  {Taniguchi}, \citenamefont {Hone}, \citenamefont {Scheurer},\ and\
  \citenamefont {Li}}]{DWPaper}%
  \BibitemOpen
  \bibfield  {author} {\bibinfo {author} {\bibfnamefont {P.}~\bibnamefont
  {Siriviboon}}, \bibinfo {author} {\bibfnamefont {J.-X.}\ \bibnamefont {Lin}},
  \bibinfo {author} {\bibfnamefont {H.~D.}\ \bibnamefont {Scammell}}, \bibinfo
  {author} {\bibfnamefont {S.}~\bibnamefont {Liu}}, \bibinfo {author}
  {\bibfnamefont {D.}~\bibnamefont {Rhodes}}, \bibinfo {author} {\bibfnamefont
  {K.}~\bibnamefont {Watanabe}}, \bibinfo {author} {\bibfnamefont
  {T.}~\bibnamefont {Taniguchi}}, \bibinfo {author} {\bibfnamefont
  {J.}~\bibnamefont {Hone}}, \bibinfo {author} {\bibfnamefont {M.~S.}\
  \bibnamefont {Scheurer}}, \ and\ \bibinfo {author} {\bibfnamefont
  {J.}~\bibnamefont {Li}},\ }\href@noop {} {\bibfield  {journal} {\bibinfo
  {journal} {arXiv e-prints}\ } (\bibinfo {year} {2021})},\ \Eprint
  {http://arxiv.org/abs/2112.07127} {arXiv:2112.07127} \BibitemShut {NoStop}%
\bibitem [{\citenamefont {{Zinkl}}\ \emph {et~al.}(2021)\citenamefont
  {{Zinkl}}, \citenamefont {{Hamamoto}},\ and\ \citenamefont
  {{Sigrist}}}]{2021arXiv211105340Z}%
  \BibitemOpen
  \bibfield  {author} {\bibinfo {author} {\bibfnamefont {B.}~\bibnamefont
  {{Zinkl}}}, \bibinfo {author} {\bibfnamefont {K.}~\bibnamefont {{Hamamoto}}},
  \ and\ \bibinfo {author} {\bibfnamefont {M.}~\bibnamefont {{Sigrist}}},\
  }\href@noop {} {\bibfield  {journal} {\bibinfo  {journal} {arXiv e-prints}\ }
  (\bibinfo {year} {2021})},\ \Eprint {http://arxiv.org/abs/2111.05340}
  {arXiv:2111.05340 [cond-mat.supr-con]} \BibitemShut {NoStop}%
\bibitem [{\citenamefont {Scheurer}(2016)}]{PhysRevB.93.174509}%
  \BibitemOpen
  \bibfield  {author} {\bibinfo {author} {\bibfnamefont {M.~S.}\ \bibnamefont
  {Scheurer}},\ }\href {\doibase 10.1103/PhysRevB.93.174509} {\bibfield
  {journal} {\bibinfo  {journal} {Phys. Rev. B}\ }\textbf {\bibinfo {volume}
  {93}},\ \bibinfo {pages} {174509} (\bibinfo {year} {2016})}\BibitemShut
  {NoStop}%
\bibitem [{\citenamefont {Liu}\ \emph {et~al.}(2021{\natexlab{a}})\citenamefont
  {Liu}, \citenamefont {Wang}, \citenamefont {Watanabe}, \citenamefont
  {Taniguchi}, \citenamefont {Vafek},\ and\ \citenamefont {Li}}]{Liu2021DtBLG}%
  \BibitemOpen
  \bibfield  {author} {\bibinfo {author} {\bibfnamefont {X.}~\bibnamefont
  {Liu}}, \bibinfo {author} {\bibfnamefont {Z.}~\bibnamefont {Wang}}, \bibinfo
  {author} {\bibfnamefont {K.}~\bibnamefont {Watanabe}}, \bibinfo {author}
  {\bibfnamefont {T.}~\bibnamefont {Taniguchi}}, \bibinfo {author}
  {\bibfnamefont {O.}~\bibnamefont {Vafek}}, \ and\ \bibinfo {author}
  {\bibfnamefont {J.}~\bibnamefont {Li}},\ }\href@noop {} {\bibfield  {journal}
  {\bibinfo  {journal} {Science}\ }\textbf {\bibinfo {volume} {371}},\ \bibinfo
  {pages} {1261} (\bibinfo {year} {2021}{\natexlab{a}})}\BibitemShut {NoStop}%
\bibitem [{\citenamefont {Liu}\ \emph {et~al.}(2021{\natexlab{b}})\citenamefont
  {Liu}, \citenamefont {Watanabe}, \citenamefont {Taniguchi},\ and\
  \citenamefont {Li}}]{Liu2021DtTLG}%
  \BibitemOpen
  \bibfield  {author} {\bibinfo {author} {\bibfnamefont {X.}~\bibnamefont
  {Liu}}, \bibinfo {author} {\bibfnamefont {K.}~\bibnamefont {Watanabe}},
  \bibinfo {author} {\bibfnamefont {T.}~\bibnamefont {Taniguchi}}, \ and\
  \bibinfo {author} {\bibfnamefont {J.}~\bibnamefont {Li}},\ }\href@noop {}
  {\bibfield  {journal} {\bibinfo  {journal} {arXiv preprint arXiv:2108.03338}\
  } (\bibinfo {year} {2021}{\natexlab{b}})}\BibitemShut {NoStop}%
\bibitem [{\citenamefont {Hao}\ \emph {et~al.}(2021)\citenamefont {Hao},
  \citenamefont {Zimmerman}, \citenamefont {Ledwith}, \citenamefont {Khalaf},
  \citenamefont {Najafabadi}, \citenamefont {Watanabe}, \citenamefont
  {Taniguchi}, \citenamefont {Vishwanath},\ and\ \citenamefont
  {Kim}}]{Hao2021tTLG}%
  \BibitemOpen
  \bibfield  {author} {\bibinfo {author} {\bibfnamefont {Z.}~\bibnamefont
  {Hao}}, \bibinfo {author} {\bibfnamefont {A.}~\bibnamefont {Zimmerman}},
  \bibinfo {author} {\bibfnamefont {P.}~\bibnamefont {Ledwith}}, \bibinfo
  {author} {\bibfnamefont {E.}~\bibnamefont {Khalaf}}, \bibinfo {author}
  {\bibfnamefont {D.~H.}\ \bibnamefont {Najafabadi}}, \bibinfo {author}
  {\bibfnamefont {K.}~\bibnamefont {Watanabe}}, \bibinfo {author}
  {\bibfnamefont {T.}~\bibnamefont {Taniguchi}}, \bibinfo {author}
  {\bibfnamefont {A.}~\bibnamefont {Vishwanath}}, \ and\ \bibinfo {author}
  {\bibfnamefont {P.}~\bibnamefont {Kim}},\ }\href@noop {} {\bibfield
  {journal} {\bibinfo  {journal} {Science}\ }\textbf {\bibinfo {volume}
  {371}},\ \bibinfo {pages} {1133} (\bibinfo {year} {2021})}\BibitemShut
  {NoStop}%
\bibitem [{\citenamefont {{Christos}}\ \emph {et~al.}(2021)\citenamefont
  {{Christos}}, \citenamefont {{Sachdev}},\ and\ \citenamefont
  {{Scheurer}}}]{2021arXiv210602063C}%
  \BibitemOpen
  \bibfield  {author} {\bibinfo {author} {\bibfnamefont {M.}~\bibnamefont
  {{Christos}}}, \bibinfo {author} {\bibfnamefont {S.}~\bibnamefont
  {{Sachdev}}}, \ and\ \bibinfo {author} {\bibfnamefont {M.~S.}\ \bibnamefont
  {{Scheurer}}},\ }\href@noop {} {\bibfield  {journal} {\bibinfo  {journal}
  {arXiv e-prints}\ } (\bibinfo {year} {2021})},\ \Eprint
  {http://arxiv.org/abs/2106.02063} {arXiv:2106.02063 [cond-mat.str-el]}
  \BibitemShut {NoStop}%
\bibitem [{\citenamefont {Agterberg}\ and\ \citenamefont
  {Kaur}(2007)}]{PhysRevB.75.064511}%
  \BibitemOpen
  \bibfield  {author} {\bibinfo {author} {\bibfnamefont {D.~F.}\ \bibnamefont
  {Agterberg}}\ and\ \bibinfo {author} {\bibfnamefont {R.~P.}\ \bibnamefont
  {Kaur}},\ }\href {\doibase 10.1103/PhysRevB.75.064511} {\bibfield  {journal}
  {\bibinfo  {journal} {Phys. Rev. B}\ }\textbf {\bibinfo {volume} {75}},\
  \bibinfo {pages} {064511} (\bibinfo {year} {2007})}\BibitemShut {NoStop}%
\bibitem [{\citenamefont {Dimitrova}\ and\ \citenamefont
  {Feigel'man}(2007)}]{PhysRevB.76.014522}%
  \BibitemOpen
  \bibfield  {author} {\bibinfo {author} {\bibfnamefont {O.}~\bibnamefont
  {Dimitrova}}\ and\ \bibinfo {author} {\bibfnamefont {M.~V.}\ \bibnamefont
  {Feigel'man}},\ }\href {\doibase 10.1103/PhysRevB.76.014522} {\bibfield
  {journal} {\bibinfo  {journal} {Phys. Rev. B}\ }\textbf {\bibinfo {volume}
  {76}},\ \bibinfo {pages} {014522} (\bibinfo {year} {2007})}\BibitemShut
  {NoStop}%
\bibitem [{\citenamefont {Scheurer}\ and\ \citenamefont
  {Samajdar}(2020)}]{PhysRevResearch.2.033062}%
  \BibitemOpen
  \bibfield  {author} {\bibinfo {author} {\bibfnamefont {M.~S.}\ \bibnamefont
  {Scheurer}}\ and\ \bibinfo {author} {\bibfnamefont {R.}~\bibnamefont
  {Samajdar}},\ }\href {\doibase 10.1103/PhysRevResearch.2.033062} {\bibfield
  {journal} {\bibinfo  {journal} {Phys. Rev. Research}\ }\textbf {\bibinfo
  {volume} {2}},\ \bibinfo {pages} {033062} (\bibinfo {year}
  {2020})}\BibitemShut {NoStop}%
\bibitem [{\citenamefont {{Naimer}}\ \emph {et~al.}(2021)\citenamefont
  {{Naimer}}, \citenamefont {{Zollner}}, \citenamefont {{Gmitra}},\ and\
  \citenamefont {{Fabian}}}]{2021arXiv210806126N}%
  \BibitemOpen
  \bibfield  {author} {\bibinfo {author} {\bibfnamefont {T.}~\bibnamefont
  {{Naimer}}}, \bibinfo {author} {\bibfnamefont {K.}~\bibnamefont {{Zollner}}},
  \bibinfo {author} {\bibfnamefont {M.}~\bibnamefont {{Gmitra}}}, \ and\
  \bibinfo {author} {\bibfnamefont {J.}~\bibnamefont {{Fabian}}},\ }\href@noop
  {} {\bibfield  {journal} {\bibinfo  {journal} {arXiv e-prints}\ } (\bibinfo
  {year} {2021})},\ \Eprint {http://arxiv.org/abs/2108.06126} {arXiv:2108.06126
  [cond-mat.mes-hall]} \BibitemShut {NoStop}%
\bibitem [{\citenamefont {Cao}\ \emph {et~al.}(2020)\citenamefont {Cao},
  \citenamefont {Rodan-Legrain}, \citenamefont {Park}, \citenamefont {Yuan},
  \citenamefont {Watanabe}, \citenamefont {Taniguchi}, \citenamefont
  {Fernandes}, \citenamefont {Fu},\ and\ \citenamefont
  {Jarillo-Herrero}}]{Cao2020nematicity}%
  \BibitemOpen
  \bibfield  {author} {\bibinfo {author} {\bibfnamefont {Y.}~\bibnamefont
  {Cao}}, \bibinfo {author} {\bibfnamefont {D.}~\bibnamefont {Rodan-Legrain}},
  \bibinfo {author} {\bibfnamefont {J.~M.}\ \bibnamefont {Park}}, \bibinfo
  {author} {\bibfnamefont {F.~N.}\ \bibnamefont {Yuan}}, \bibinfo {author}
  {\bibfnamefont {K.}~\bibnamefont {Watanabe}}, \bibinfo {author}
  {\bibfnamefont {T.}~\bibnamefont {Taniguchi}}, \bibinfo {author}
  {\bibfnamefont {R.~M.}\ \bibnamefont {Fernandes}}, \bibinfo {author}
  {\bibfnamefont {L.}~\bibnamefont {Fu}}, \ and\ \bibinfo {author}
  {\bibfnamefont {P.}~\bibnamefont {Jarillo-Herrero}},\ }\href@noop {}
  {\bibfield  {journal} {\bibinfo  {journal} {arXiv preprint arXiv:2004.04148}\
  } (\bibinfo {year} {2020})}\BibitemShut {NoStop}%
\bibitem [{\citenamefont {Lin}\ \emph {et~al.}(2022)\citenamefont {Lin},
  \citenamefont {Zhang}, \citenamefont {Morissette}, \citenamefont {Wang},
  \citenamefont {Liu}, \citenamefont {Rhodes}, \citenamefont {Watanabe},
  \citenamefont {Taniguchi}, \citenamefont {Hone},\ and\ \citenamefont
  {Li}}]{Lin2021SOC}%
  \BibitemOpen
  \bibfield  {author} {\bibinfo {author} {\bibfnamefont {J.-X.}\ \bibnamefont
  {Lin}}, \bibinfo {author} {\bibfnamefont {Y.-H.}\ \bibnamefont {Zhang}},
  \bibinfo {author} {\bibfnamefont {E.}~\bibnamefont {Morissette}}, \bibinfo
  {author} {\bibfnamefont {Z.}~\bibnamefont {Wang}}, \bibinfo {author}
  {\bibfnamefont {S.}~\bibnamefont {Liu}}, \bibinfo {author} {\bibfnamefont
  {D.}~\bibnamefont {Rhodes}}, \bibinfo {author} {\bibfnamefont
  {K.}~\bibnamefont {Watanabe}}, \bibinfo {author} {\bibfnamefont
  {T.}~\bibnamefont {Taniguchi}}, \bibinfo {author} {\bibfnamefont
  {J.}~\bibnamefont {Hone}}, \ and\ \bibinfo {author} {\bibfnamefont
  {J.}~\bibnamefont {Li}},\ }\href@noop {} {\bibfield  {journal} {\bibinfo
  {journal} {Science}\ }\textbf {\bibinfo {volume} {375}},\ \bibinfo {pages}
  {437} (\bibinfo {year} {2022})}\BibitemShut {NoStop}%
\bibitem [{\citenamefont {Serlin}\ \emph {et~al.}(2019)\citenamefont {Serlin},
  \citenamefont {Tschirhart}, \citenamefont {Polshyn}, \citenamefont {Zhang},
  \citenamefont {Zhu}, \citenamefont {Watanabe}, \citenamefont {Taniguchi},
  \citenamefont {Balents},\ and\ \citenamefont {Young}}]{Serlin2019}%
  \BibitemOpen
  \bibfield  {author} {\bibinfo {author} {\bibfnamefont {M.}~\bibnamefont
  {Serlin}}, \bibinfo {author} {\bibfnamefont {C.}~\bibnamefont {Tschirhart}},
  \bibinfo {author} {\bibfnamefont {H.}~\bibnamefont {Polshyn}}, \bibinfo
  {author} {\bibfnamefont {Y.}~\bibnamefont {Zhang}}, \bibinfo {author}
  {\bibfnamefont {J.}~\bibnamefont {Zhu}}, \bibinfo {author} {\bibfnamefont
  {K.}~\bibnamefont {Watanabe}}, \bibinfo {author} {\bibfnamefont
  {T.}~\bibnamefont {Taniguchi}}, \bibinfo {author} {\bibfnamefont
  {L.}~\bibnamefont {Balents}}, \ and\ \bibinfo {author} {\bibfnamefont
  {A.}~\bibnamefont {Young}},\ }\href@noop {} {\bibfield  {journal} {\bibinfo
  {journal} {arXiv preprint arXiv:1907.00261}\ } (\bibinfo {year}
  {2019})}\BibitemShut {NoStop}%
\bibitem [{\citenamefont {Chandrasekhar}(1962)}]{Chandrasekhar1962Pauli}%
  \BibitemOpen
  \bibfield  {author} {\bibinfo {author} {\bibfnamefont {B.}~\bibnamefont
  {Chandrasekhar}},\ }\href@noop {} {\bibfield  {journal} {\bibinfo  {journal}
  {Applied Physics Letters}\ }\textbf {\bibinfo {volume} {1}},\ \bibinfo
  {pages} {7} (\bibinfo {year} {1962})}\BibitemShut {NoStop}%
\bibitem [{\citenamefont {Clogston}(1962)}]{Clogston1962Pauli}%
  \BibitemOpen
  \bibfield  {author} {\bibinfo {author} {\bibfnamefont {A.~M.}\ \bibnamefont
  {Clogston}},\ }\href@noop {} {\bibfield  {journal} {\bibinfo  {journal}
  {Physical Review Letters}\ }\textbf {\bibinfo {volume} {9}},\ \bibinfo
  {pages} {266} (\bibinfo {year} {1962})}\BibitemShut {NoStop}%
\end{thebibliography}%

\newpage

\newpage
\clearpage

\pagebreak
\begin{widetext}
\section{Supplementary Materials}

\begin{center}
\textbf{\large Zero-field superconducting diode effect in twisted trilayer graphene}\\
\vspace{10pt}

Jiang-Xiazi Lin, Phum Siriviboon,  Harley D. Scammell, Song Liu, Daniel Rhodes, K. Watanabe, T. Taniguchi, James Hone, Mathias S. Scheurer, and J.I.A. Li$^{\dag}$

\vspace{10pt}
$^{\dag}$ Corresponding author. Email: jia$\_$li@brown.edu
\end{center}

\noindent\textbf{This PDF file includes:}

\noindent{Supplementary Text}

\noindent{Materials and Methods}

\noindent{Figs. S1 to S12}

\renewcommand{\vec}[1]{\boldsymbol{#1}}

\renewcommand{\thefigure}{S\arabic{figure}}
\def\theequation{S\arabic{equation}}
\def\thetable{S\Roman{table}}
\setcounter{figure}{0}
\setcounter{equation}{0}

\newpage

\subsection{SI 1: Symmetry analysis for superconducting diode effect}

Let us begin with a discussion of general constraints on the superconducting diode effect based on symmetry. Denoting the magnitude of the maximum current density the superconductor can support along the in-plane direction $\hat{n}$ by $J_c(\hat{n})$, the supercurrent diode effect is defined by 
\begin{equation}
    \exists \hat{n}:\quad \delta J_c(\hat{n}) := J_c(\hat{n}) - J_c(-\hat{n})\neq 0. 
\end{equation}
Since the current direction is odd under time-reversal and under two-fold out-of-plane rotation or three-dimensional inversion symmetry, the presence of at least one of these symmetries implies that
\begin{equation}
    J_c(\hat{n}) = J_c(-\hat{n}) \label{SymmetryConstraint}
\end{equation}
and no diode effect can be present. In turn, the observation of a finite diode effect in our sample unambiguously demonstrates that all three of these symmetries must be broken in the superconducting state.

There are two qualitatively different ways for this to happen, which are in principle consistent with our observations: (i) all three of these symmetries are already broken in the normal state out of which superconductivity emerges or (ii) at least one of these symmetries is broken spontaneously when (the magnitude) of the superconducting order parameter becomes finite. As already discussed in the main text, we believe that option (ii) is less natural since the underlying pairing must be unconventional; we therefore focus on (i) in the following and refer to \cite{DiodeTheoryPaper} for more details on (ii).

\begin{table*}[b]
\begin{center}
\caption{Action of the point symmetries on the microscopic field operators ($\psi_{\vec{k}}$) of a continuum-model description of twisted trilayer graphene, using the same notation as in \cite{2021arXiv210602063C,DWPaper,DiodeTheoryPaper}. For convenience of the reader, we also list redundant symmetries. Here $\beta$ is a parameter that describes strain-induced $C_{3z}$-symmetry breaking and $\lambda_{\text{R}}$ ($\lambda_{\text{I}}$) denotes Rashba (Ising) spin-orbit coupling.}
\label{ActionOfSymmetries}
\begin{ruledtabular}
 \begin{tabular} {cccc} 
Symmetry $S$ & unitary  & $S\psi_{\vec{k};\ell,\vec{G}}S^\dagger$ & condition \\ \hline
SO(3)$_{s}$ & \cmark    & $e^{i\vec{\varphi}\cdot \vec{s}}  \psi_{\vec{k};\ell,\vec{G}}$  & $\lambda_{\text{R}}=\lambda_{\text{I}}=0$  \\ 
SO(2)$_{s}$ & \cmark    & $e^{i \varphi s_z}  \psi_{\vec{k};\ell,\vec{G}}$  & $\lambda_{\text{R}}=0$  \\ \hline
$C_{3z}$ & \cmark    & $e^{i\frac{2\pi}{3} \rho_3\eta_3}  \psi_{C_{3z}\vec{k};\ell,C_{3z}\vec{G}}$  & $\lambda_{\text{R}}=\beta=0$  \\
$C^s_{3z}$ & \cmark    & $e^{i\frac{2\pi}{3} (\rho_3\eta_3 + s_3)}  \psi_{C_{3z}\vec{k};\ell,C_{3z}\vec{G}}$  & $\beta=0$  \\ \hline
$C_{2z}$ & \cmark   & $\eta_1 \rho_1 \psi_{-\vec{k};\ell,-\vec{G}}$  & $\lambda_{\text{R}}=\lambda_{\text{I}}=0$  \\
$C^s_{2z}$ & \cmark   & $s_3\eta_1 \rho_1 \psi_{-\vec{k};\ell,-\vec{G}}$  & $\lambda_{\text{I}}=0$  \\
$C^{s'}_{2z}=C^s_{2z} i  s_{2,1}$ & \cmark   & $s_{1,2}\eta_1 \rho_1 \psi_{-\vec{k};\ell,-\vec{G}}$  & $\lambda_{\text{R}}=0$  \\ 
$\sigma_h$ & \cmark   & $ (1,1,-1)_\ell \psi_{\vec{k};\ell,\vec{G}}$   & $D_0=\lambda_{\text{R}}=\lambda_{\text{I}}=0$  \\
$I = C_{2z}\sigma_h$ & \cmark    & $ \eta_1\rho_1(1,1,-1)_\ell \psi_{-\vec{k};\ell,\vec{G}}$   & $D_0=\lambda_{\text{R}}=\lambda_{\text{I}}=0$  \\ \hline
$\Theta$ & \xmark    & $\eta_1 \psi_{-\vec{k};\ell,-\vec{G}}$  & $\lambda_{\text{R}}=\lambda_{\text{I}}=0$  \\
$\Theta^s$ & \xmark   & $is_2\eta_1 \psi_{-\vec{k};\ell,-\vec{G}}$  & ---  \\

 \end{tabular}
\end{ruledtabular}
\end{center}
\end{table*}

In \tableref{ActionOfSymmetries}, we list the point symmetries of mirror-symmetric twisted trilayer graphene, as a function of whether a displacement field $D_0$ is applied, Rashba, $\lambda_{\text{R}}$, or Ising, $\lambda_{\text{I}}$, spin-orbit coupling is present, and whether there is strain, $\beta$, in the samples. For completeness, we also list their representation on the continuum-model field operators $(\psi_{\vec{k};\ell,\vec{G}})_{\rho,\eta,s}$ with $s$ denoting the spin, $\rho$ the sublattice, and $\eta$ the valley quantum number (and the associated Pauli matrices are denoted by $s_j$, $\rho_j$, and $\eta_j$); $\ell$ labels the two mirror even $\ell=1,2$  and the mirror-odd, $\ell=3$, layer wavefunctions (see \cite{2021arXiv210602063C,DWPaper,DiodeTheoryPaper} where the same conventions are used), and $\vec{G}$ are vectors of the reciprocal moir\'e lattice. Note that we list both spinless forms of the symmetries (e.g., $C_{2z}$), which are only possible in the absence of spin-orbit coupling, and spinfull versions ($C_{2z}^s$ or $C_{2z}^{s'}$), where the spatial transformations are combined with appropriate spin transformations. Importantly, the constraint in \equref{SymmetryConstraint} follows when any one of the spinfull or spinless forms of these symmetries, i.e., any of $\Theta$, $\Theta^s$, $C_{2z}$, $C_{2z}^s$, and $C_{2z}^{s'}$, are present. 

While spin-orbit coupling breaks all two-fold rotational symmetries, $\Theta^s$ is always preserved (which also holds when further spin-orbit terms, like Kane-Mele-type spin-orbit coupling, or a sublattice imbalance term, $m\rho_z$, that could be induced by WSe$_2$, are taken into account). In scenario (i) introduced above, the normal state above the superconducting transition temperature must therefore exhibit spontaneously broken time-reversal symmetry, induced by electronic interactions. Furthermore, the associated particle-hole order parameter must coexist with superconductivity. We next discuss a few natural microscopic candidate forms for this symmetry-breaking order.

\SIsubsec{Different normal state orders} The first type of order parameter breaking time-reversal symmetry is spin magnetism. The associated three-component order parameter, $\vec{S}$, couples to the electrons according to
\begin{equation}
    \Delta H_S = \sum_{\vec{k}\in\text{MBZ}}\sum_{\vec{G}\in\text{RML}}\sum_{\ell=1}^3 g_{\ell,\vec{G}}(\vec{k}) \, \psi^\dagger_{\vec{k};\ell,\vec{G}} \vec{s} \psi^\pdagger_{\vec{k};\ell,\vec{G}} \cdot \vec{S}, \label{SpinMagnetismOrderParameter}
\end{equation}
where MBZ (RML) denotes the moir\'e Brillouin zone (reciprocal moir\'e lattice), $\vec{s}=(s_x,s_y,s_y)$ are Pauli matrices in spin space, and $g_{\ell,\vec{G}}\in\mathbbm{R}$ with $g_{\ell,\vec{G}}(\vec{k})$ only constrained by $C_{2z}$ [$g_{\ell,\vec{G}}(\vec{k})=g_{\ell,-\vec{G}}(-\vec{k})$] and $C_{3z}$ [$g_{\ell,\vec{G}}(\vec{k})=g_{\ell,C_{3z}\vec{G}}(C_{3z}\vec{k})$]. Among the symmetries in \tableref{ActionOfSymmetries} of the model without spin-orbit coupling and strain, $\lambda_{\text{I}}=\lambda_{\text{R}}=\beta=0$, non-zero $\vec{S}$ in \equref{SpinMagnetismOrderParameter} only breaks $\text{SO}(3)_s$ (down to the group of residual spin-rotations along the direction of $\vec{S}$). In the presence of any finite subset of $\lambda_{\text{I}}$, $\lambda_{\text{R}}$, $\beta$, additional symmetries are broken, with important consequences for the diode effect, as detailed below. 

Note that in the presence of $\lambda_{\text{I}}$ or $\lambda_{\text{R}}$, the spin-rotation symmetry SO(3)$_s$ is broken (down to SO(2)$_s$ or completely). Therefore, out-of-plane spin magnetization, $\vec{S} = (0,0,S_z)$, and in-plane magnetization, $\vec{S} = (S_x,S_y,0)$, are symmetry inequivalent and, hence, constitute physically distinct order parameters; we will thus analyze the diode effect for these two types of orientations of $\vec{S}$ separately below.

Before discussing that, though, let us introduce the second class of time-reversal-symmetry-breaking order, which is spontaneous valley polarization, $V_z$, with coupling
\begin{equation}
    \Delta H_V = \sum_{\vec{k}\in\text{MBZ}}\sum_{\vec{G}\in\text{RML}}\sum_{\ell=1}^3 \tilde{g}_{\ell,\vec{G}}(\vec{k}) \, \psi^\dagger_{\vec{k};\ell,\vec{G}} \eta_z \psi^\pdagger_{\vec{k};\ell,\vec{G}} \cdot V_z, \label{ValleyPolOrderParameter}
\end{equation}
where $\eta_z$ is the third Pauli matrix in valley space and $\tilde{g}_{\ell,\vec{G}}(\vec{k})$ is constrained exactly in the same way as $g$ above. A finite valley polarization, $V_z\neq 0$, breaks both time-reversal and two-fold-rotation symmetry (along with inversion $I$) at the same time, as immediately follows from the representations in \tableref{ActionOfSymmetries}.

\SIsubsec{Consequences for diode effect} Using the symmetry criteria defined above, it is straightforward to determine whether a diode effect is allowed for the three different types of normal state orders---in-plane ($S_{x,y}$), out-of-plane ($S_z$) spin magnetization, and valley ($V_z$) polarization---in the presence or absence of spin-orbit coupling and strain. The result is summarized in the third column of \tableref{DiodeEffect}. The findings imply, for instance, that the diode effect in case of out-of-plane spin magnetism has to be proportional to both $S_z$ and the Ising-spin-orbit coupling strength, $\delta J(\hat{n}) \propto S_z\lambda_{\text{I}}$; this simply follows from the fact that $S_z$ alone (or combined with Rashba spin-orbit coupling or $\beta$) does not break $C_{2z}^s$ and, therefore, does not allow for a diode effect. Similarly, we have $\delta J(\hat{n}) \propto S_{x,y}\lambda_{\text{R}}$ for in-plane spin magnetism, as Rashba spin-orbit coupling is required to break $C_{2z}^{s'}$. In case of valley polarization, this is different as it breaks all two-fold rotational symmetries (and time-reversal) even without any spin-orbit coupling and, hence, $\delta J(\hat{n}) \propto V_z \neq 0$ even when $\lambda_{\text{I}}=\lambda_{\text{R}}=\beta=0$. 

Note that $V_z$ and $S_z$ are, from a pure symmetry perspective, equivalent as long as $\lambda_{\text{I}} \neq 0$, while $S_{x,y}$ is still distinct; one important qualitative difference between out-of-plane spin polarization, $S_z$, or valley polarization, $V_z$, being responsible for the anomalous diode effect as compared to in-plane spin polarization, $S_{x,y}$, is that the diode effect for the latter will not be $C_{3z}$ symmetric: as can be seen in the third and fourth column of \tableref{DiodeEffect}, whenever we have a diode effect for $S_{x,y}$, which happens as long as $\lambda_{\text{R}} \neq 0$, the critical current will not be $C_{3z}$ symmetric, $J_c(\hat{n}) \neq J_c(C_{3z}\hat{n})$. This means that the current asymmetry, $\delta J_c(\hat{n}) := J_c(\hat{n}) - J_c(-\hat{n})$, is constrained to have at least two zeros when $\hat{n}$ rotates by $2\pi$. This is different for $S_z$ or $V_z$; as long as strain is negligible, $\beta=0$, it holds $J_c(\hat{n}) = J_c(C_{3z}\hat{n})$ and consequently $\delta J_c(\hat{n})$ must have at least six zeros. Unless strain is sufficiently large, the number of sign changes will not be affected by $\beta\neq0$. 

By the same token, the response of the superconductor to an applied in-plane magnetic field $\vec{B} = (B_x,B_y,0) = |\vec{B}| \hat{b}$, e.g., as reflected in its critical temperature, will display different angular dependence: $T_c$ will be a (an approximately) $C_{3z}$ symmetric function as long as $\beta=0$ ($\beta$ is small) for $V_z$ and $S_z$, while it will not be invariant under a $C_{3z}$ rotation of the magnetic field, see fifth column in \tableref{DiodeEffect}, if $S_{x,y}$ causes the diode effect. This could be used to further check the order parameter underlying the diode effect in future experiments.

\SIsubsec{Field training} Another aspect that can be used to distinguish between the different candidate orders is whether they can be trained by in- and out-of-plane magnetic fields. Formally, we classify an order parameter $O\in\{ S_{z}, S_{x,y}, V_z\}$ as trainable by a field $\mathcal{B}$ if a linear coupling $c\, O \mathcal{B}$ is allowed by the symmetries of the bare Hamiltonian (in turn depending on which of $\lambda_{\text{I}}, \lambda_{\text{R}}, \beta$ are non-zero) without the order $O$ or $\mathcal{B}$. In the last four columns of \tableref{DiodeEffect}, we list under which conditions this is possible for the three different orders and considering orbital and Zeeman coupling separately.  

Most importantly, $S_{x,y}$ can only be trained by an out-of-plane magnetic field if there is finite strain, $\beta \neq 0$, Rashba and Ising spin-orbit coupling $\lambda_{\text{R}},\lambda_{\text{I}}\neq 0$. As such, the coupling strength $c \propto \beta \lambda_{\text{R}}\lambda_{\text{I}}$ is expected to be rather weak. In particular, much smaller than the coupling strength and, hence, trainability of $S_{x,y}$ with respect to in-plane magnetic fields, which does not require any of $\lambda_{\text{I}}$, $\lambda_{\text{R}}$, $\beta$ to be non-zero (for the Zeeman part). This is the exact opposite of what is seen in experiment, where the diode effect can be efficiently trained with an out-of-plane magnetic field of order $25\,\textrm{mT}$, while it is much more difficult to train it with in-plane magnetic field (for those in-plane fields, of order $10\,\textrm{T}$, where we seem to have achieved training of the diode effect, the lower bound on the simultaneously applied out-of-plane field is also of order of $10\,\textrm{mT}$).  

Instead, the observed behavior is consistent with either $V_z$ or $S_z$ since both can be trained by an out-of-plane Zeeman field (with the only difference that $V_z$ further requires $\lambda_{\text{I}}$ which is not small) but can only be trained by an in-plane field if $\beta$ and $\lambda_{\text{R}}$ are non-zero. Taken together, our measurements indicate that either $V_z$ or $S_z$ are the most natural candidate order parameters. We reiterate that, as long as $\lambda_{\text{I}}$ is non-zero, as is the case for our system, it is not possible to distinguish between these remaining two options by symmetry; in fact, their order parameters are in the same irreducible representation of the high-temperature symmetry group such that they formally describe the same phase. Intuitively, this can be seen by noting that Ising spin-orbit coupling has the form $\lambda_{\text{I}}s_z\eta_z$ such that a finite valley polarization will induce a spin-polarization along the spin-$z$ axis, and vice versa.

\begin{table*}[tb]
\begin{center}
\caption{We list for three different candidate orders of the normal state---out-of-plane spin polarization ($S_z$), in-plane spin polarization ($S_{x,y}$), and valley polarization ($V_z$)---whether a supercurrent diode effect, $J_c(\hat{n}) \neq J_c(-\hat{n})$, is allowed, depending on which of the three parameters, Rashba, $\lambda_{\text{R}}$, and Ising, $\lambda_{\text{I}}$, spin-orbit coupling, and strain, $\beta$, are non-zero. We further indicate whether the directional dependence of the critical current shows a free-fold rotation symmetry and whether the critical temperature of the superconductor is unaffected by rotating an applied magnetic field by $C_{3z}$. Finally, the last four columns indicated whether it can be trained by an in-plane (out-of-plane) Zeeman field $B^Z_{\parallel}$ ($B^Z_{\perp}$) or orbital magnetic field $B^O_{\parallel}$ ($B^O_{\perp}$). Throughout, we assume that the superconductor does not spontaneously break any additional point symmetry. Note that all results hold for both $D_0=0$ and $D_0\neq 0$.}
\label{DiodeEffect}
\begin{ruledtabular}
 \begin{tabular} {ccccccccc} 
\multicolumn{2}{c}{normal state} &  \multicolumn{3}{c}{superconductor} & \multicolumn{4}{c}{trainable by} \\ \cline{1-2} \cline{3-5} \cline{6-9}  
parent order & non-zero terms & $J_c(\hat{n}) \neq J_c(-\hat{n})$ & $J_c(\hat{n}) = J_c(C_{3z}\hat{n})$ & $T_c(\hat{b}) = T_c(C_{3z}\hat{b})$ & $B^Z_{\perp}$ & $B^Z_{\parallel}$ & $B^O_{\perp}$ & $B^O_{\parallel}$ \\ \hline
$S_z$ & --- & \xmark & \cmark & \cmark & \cmark & \xmark & \xmark & \xmark \\
 & $\lambda_{\text{R}}$ & \xmark &  \cmark & \cmark & \cmark & \xmark & \cmark & \xmark \\
 & $\lambda_{\text{I}}$ & \cmark &  \cmark & \cmark & \cmark & \xmark & \xmark & \xmark \\
  & $\lambda_{\text{R}},\lambda_{\text{I}}$ & \cmark & \cmark & \cmark & \cmark & \xmark & \cmark & \xmark \\
  & $\lambda_{\text{R}},\lambda_{\text{I}},\beta$ & \cmark & \xmark & \xmark & \cmark & \cmark & \cmark & \cmark \\ \hline
  $S_{x,y}$ & --- & \xmark & \cmark & \xmark & \xmark & \cmark & \xmark & \xmark \\
 & $\lambda_{\text{R}}$ & \cmark & \xmark & \xmark & \xmark & \cmark & \xmark & \cmark \\
 & $\lambda_{\text{I}}$ & \xmark & \cmark & \xmark & \xmark & \cmark & \xmark & \xmark \\
  & $\lambda_{\text{R}},\lambda_{\text{I}}$ & \cmark & \xmark & \xmark & \xmark & \cmark & \xmark & \cmark \\
  & $\lambda_{\text{R}},\lambda_{\text{I}},\beta$ & \cmark & \xmark & \xmark & \cmark & \cmark & \cmark & \cmark \\ \hline
  $V_{z}$ & --- & \cmark & \cmark & \cmark & \xmark  & \xmark  & \xmark & \xmark \\
 & $\lambda_{\text{R}}$ & \cmark & \cmark & \cmark & \xmark & \xmark & \xmark & \xmark \\
 & $\lambda_{\text{I}}$ & \cmark & \cmark & \cmark & \cmark & \xmark & \xmark & \xmark \\
  & $\lambda_{\text{R}},\lambda_{\text{I}}$ & \cmark & \cmark & \cmark & \cmark & \xmark  & \cmark & \xmark \\
  & $\lambda_{\text{R}},\lambda_{\text{I}},\beta$ & \cmark & \xmark & \xmark & \cmark & \cmark & \cmark & \cmark \\
 \end{tabular}
\end{ruledtabular}
\end{center}
\end{table*}

\subsection{SI 2: Absence of diode effect from in-plane field and Pauli limit violation} 
Due to the spin-orbit coupling induced in our sample by \WSe, one might expect that the mechanism discussed in \cite{Yuan2021diodes,DaidoSCDiode,HeSCDiode} for in-plane Zeeman fields and Rashba spin-orbit coupling, motivated by experiments \cite{Ando2020diodes} in artificial metal films, should apply here as well. This mechanism is based on the observation that an in-plane magnetic field will move Rashba spin-orbit-split Fermi surfaces in opposite directions leading to finite momentum pairing \cite{PhysRevB.75.064511,PhysRevB.76.014522} and a diode effect \cite{Yuan2021diodes,DaidoSCDiode,HeSCDiode}.
However, as can be seen in \figref{fig:Pauli}d, this is not the case in our experiments; as we will detail below, we expect this to result from the rather different nature of the spin-orbit coupling in our system.

To demonstrate this, let us focus on Ising spin-orbit coupling (setting $\lambda_{\text{R}}=0$); the inclusion of Rashba spin-orbit coupling is addressed in \cite{DiodeTheoryPaper}. 
Only keeping the low-energy bands around the Fermi surface, with field operators $f^\dagger_{\vec{k},\eta,s}$ creating an electron at momentum $\vec{k}$, in valley $\eta$, and of spin $s$, an effective Hamiltonian reads as
\begin{equation}
    H_{\text{LE}} = \sum_{\vec{k},s,\eta} f^\dagger_{\vec{k},\eta,s} \left( \epsilon_{\eta\cdot\vec{k}} + \Gamma_{\text{I}}(\vec{k};\eta) (s_z)_{s,s'}\,\eta  + B_x (s_x)_{s,s'}\right) f^\pdagger_{\vec{k},\eta,s'} + \sum_{\vec{k}} \left[ f^\dagger_{\vec{k},+,s}  (\Delta_{\vec{k}})_{s,s'} f^\dagger_{-\vec{k},-,s'} + \text{H.c.} \right] + \delta H_{\text{LE}}. \label{LEHamiltonian}
\end{equation}
Here $\Delta_{\vec{k}}= \left( \Delta^s_{\vec{k}}s_0 + \vec{d}_{\vec{k}}\cdot\vec{s} \right)is_y$ is the superconducting order parameter, $B_x$ an in-plane magnetic field (along $x$ without loss of generality), and $\delta H_{\text{LE}}$ are interaction terms, which we will not further have to specify and only assume that they preserve the, for superconductivity crucial \cite{PhysRevResearch.2.033062}, enhanced SU(2)$_+ \times$SU(2)$_-$ spin symmetry. Finally, $\Gamma_{\text{I}}(\vec{k};\eta)$, constrained by
\begin{equation}
    \Gamma_{\text{I}}(\vec{k};\eta) = \Gamma_{\text{I}}(-\vec{k};-\eta) \label{TRConstraintOnGamma}
\end{equation}
due to time-reversal symmetry (or, equivalently, $C_{2z}^{s'}$), parameterizes the impact of the Ising spin-orbit coupling in the band-projected theory. For momenta $\vec{k}$ where the mirror-even and mirror-odd bands of the trilayer system are energetically well separated, we have $\Gamma_{\text{I}}(\vec{k};\eta) \approx \lambda_{\text{I}}/2$ (using the conventions of \cite{DWPaper,DiodeTheoryPaper}), since the mirror-off-diagonal matrix elements of the Ising spin-orbit coupling only have a minor impact.

To simplify the form of \equref{LEHamiltonian}, let us perform the following transformation,
\begin{equation}
    f_{\vec{k},\eta,s} \quad  \longrightarrow \quad \left(e^{-i\frac{\varphi_\eta(\vec{k})}{2}\eta s_y}\right)_{s,s'} f_{\vec{k},\eta,s'} \label{TransformationOfFields}
\end{equation}
of the field operators. Choosing $\varphi_\eta(\vec{k}) = \arctan (B_x/\Gamma_{\text{I}}(\vec{k};\eta))$, we absorb the impact of the magnetic field in \equref{LEHamiltonian} by a rescaling of the Ising spin-orbit term,
\begin{equation}
    H_{\text{LE}} \longrightarrow \sum_{\vec{k},s,\eta} f^\dagger_{\vec{k},\eta,s} \left( \epsilon_{\eta\cdot\vec{k}} + \sqrt{\Gamma^2_{\text{I}}(\vec{k};\eta) +B_x^2} \,\, (s_z)_{s,s'}\,\eta\right) f^\pdagger_{\vec{k},\eta,s'} + \sum_{\vec{k}} \left[ f^\dagger_{\vec{k},+,s}  (\Delta'_{\vec{k}})_{s,s'} f^\dagger_{-\vec{k},-,s'} + \text{H.c.} \right] + \delta H'_{\text{LE}}. 
\end{equation}
This already shows that the impact of the in-plane Zeeman field on the bandstructure and Fermi surfaces is rather different from that of the ``conventional Rashba term'' ($k_ys_x-k_xs_y$) studied in \cite{PhysRevB.75.064511,PhysRevB.76.014522,Yuan2021diodes,DaidoSCDiode,HeSCDiode}: instead of shifting Fermi surfaces in opposite directions, we can see that $B_x\neq 0$ only rescales the Ising spin-orbit coupling strength. As such, for each state at momentum $\vec{k}$ in valley $\eta$ there has to be another state with the same energy at momentum $-\vec{k}$ in valley $-\eta$ and no finite momentum pairing is expected. Furthermore, there will be no diode effect since $C_{2z}^{s'}$ is still a symmetry. In fact, from a symmetry point of view, this conclusion is equivalent to the absence of a diode effect documented in Table~\ref{DiodeEffect} for $S_{x,y}$ order and $\lambda_R=0$, since $S_x$ in \equref{SpinMagnetismOrderParameter} and $B_x$ in \equref{LEHamiltonian} transform the same way under all symmetries of the system. 

Note, however, that $B_x\neq 0$ does affect the spin-polarization of the Bloch states at the Fermi surface, an effect that becomes significant if $B_x$ is of order of $\lambda_{\text{I}}/2$. The accurate value of $\lambda_{\text{I}}$ for our sample is not known, but we expect it to be in the range $\lambda_{\text{I}}/2 \approx 0.5-5 \,\textrm{meV}$ \cite{2021arXiv210806126N,DWPaper}. This corresponds to a magnetic field strength of order of or larger than $8 \,\textrm{T}$. From \equref{TransformationOfFields} and using \equref{TRConstraintOnGamma}, it follows that turning on $B_x$ can be understood as a, in general $\vec{k}$ dependent, SU(2)$_+ \times$SU(2)$_- \simeq $ SO(4) transformation of the singlet and triplet component,
\begin{equation}
    \begin{pmatrix} \Delta^s_{\vec{k}} \\ d^x_{\vec{k}} \\ d^y_{\vec{k}} \\ d^z_{\vec{k}} \end{pmatrix}  \quad \longrightarrow \quad \begin{pmatrix} \cos \varphi_+(\vec{k}) & 0 & i\sin \varphi_+(\vec{k}) & 0 \\ 0 & 1 & 0 & 0 \\
    i \sin \varphi_+(\vec{k}) & 0 & \cos \varphi_+(\vec{k}) & 0 \\
    0 & 0 & 0 & 1
    \end{pmatrix} \begin{pmatrix} \Delta^s_{\vec{k}} \\ d^x_{\vec{k}} \\ d^y_{\vec{k}} \\ d^z_{\vec{k}} \end{pmatrix}.
\end{equation}
This shows that, for instance, a finite singlet component $\Delta^s_{\vec{k}}$ at $B_x=0$ will smoothly ``rotate into'' a $d^y_{\vec{k}}$ component with relative phase $i$ when $B_x$ is turned on. Note that even a  SU(2)$_+ \times$SU(2)$_-$ symmetric interaction $\delta H'_{\text{LE}}$ will not be invariant under a $\vec{k}$-dependent transformation (\ref{TransformationOfFields}), leading to $\delta H'_{\text{LE}} \neq \delta H_{\text{LE}}$. If, however, for the for superconductivity relevant parts of the MBZ $\Gamma_{\text{I}}(\vec{k};\eta) \approx \text{const.}$ is a good approximation, we will have $\varphi_\eta(\vec{k}) \approx \varphi$ and, thus, $\delta H'_{\text{LE}} \approx \delta H_{\text{LE}}$ for any SU(2)$_+ \times$SU(2)$_-$ symmetric interaction. Apart from the likely irrelevant impact of the rescaling $\lambda_{\text{R}} \rightarrow \sqrt{\lambda_{\text{R}}^2+(2B_x)^2}$, the critical temperature of superconductivity is (approximately) not affected by an in-plane Zeeman field, which can explain the Pauli-limit violation seen in \figref{fig:Pauli}d. Eventually, superconductivity will be suppressed by the in-plane orbital coupling associated with Peierls phases in the intralayer hopping \cite{Cao2020nematicity}.

Note that in-plane orbital coupling alone cannot induce a diode effect either: as can be seen in \tableref{ActionOfSymmetries}, the system exhibits the mirror symmetry $\sigma_h$ as long as $D_0=0$ and in the absence of spin-orbit coupling. Since the in-plane orbital coupling is odd under $\sigma_h$, it is even under $\Theta \sigma_h$ which thus remain unbroken. Since the current, in turn, is odd under $\Theta \sigma_h$, \equref{SymmetryConstraint} holds, proving that an in-plane orbital coupling cannot induce a diode effect either in twisted trilayer graphene (as opposed to twisted bilayer graphene where this should be possible by symmetry \cite{Yuan2021diodes}).

\subsection{SI 3: Superconductivity in tTLG/\WSe\ heterostructures with different twist angle}

\begin{table}[h]
\begin{center}
\caption{We list material properties of five mirror-symmetric twisted trilayer graphene samples investigated in this work, such as twist angle and proximity with a \WSe\ crystal ~\cite{DWPaper}. Superconductivity is observed in all the samples except sample E with a twist angle of $\theta = 1.38 ^{\circ}$.  In sample A, C and E, a few-layer \WSe\ crystal is stacked  on top of tTLG, whereas tTLG in sample B and D are fully encapculated by hexagonal boron nitride crystals and the heterestructures do not contain \WSe.   
}
\label{TableSample}
\begin{ruledtabular}
 \begin{tabular} {ccccc} 
 \\
\multicolumn{1}{c}{Sample} & \multicolumn{1}{c}{$\theta$} & \multicolumn{1}{c}{\WSe} & \multicolumn{1}{c}{$T_c$ (K)} & \multicolumn{1}{c}{zero-field superconducting diode effect} \\  
\\
\hline
\\
A & $1.25^{\circ}\pm0.01^{\circ}$ & \cmark &  $ 1$  & \cmark   \\
\\
 \hline 
\\
B & $1.31^{\circ}\pm0.01^{\circ}$ & \xmark &  $0.9$ & \cmark  \\
\\
 \hline
\\
C & $1.47^{\circ}\pm0.02^{\circ}$ & \cmark &  $1.4$  & \xmark   \\
\\
  \hline
\\
D & $1.50^{\circ}\pm0.02^{\circ}$ & \xmark &  $1.9$  & \xmark   \\
\\
    \hline
\\
E & $1.38^{\circ}\pm0.01^{\circ}$ & \cmark &  $0$ (no superconductivity)  & -  \\
\\
 \end{tabular}
\end{ruledtabular}
\end{center}
\end{table}

\begin{table}[h]
\begin{center}
\caption{We summarize the main observations of different twist angle regimes. The twist angle dependence of flavor polarization and linear-in-T behavior is discussed in Ref. ~\cite{DWPaper}.
}
\label{TableCompare}
\begin{ruledtabular}
 \begin{tabular}{ccc} 
 \\
\multicolumn{1}{c}{} & \multicolumn{1}{c}{Near the magic angle} & \multicolumn{1}{c}{small twist angle regime}   \\  
\\
\hline
\\
Superconductivity & \cmark & \cmark    \\
\\
 \hline 
\\
Flavor polarization & \cmark & \xmark   \\
\\
 \hline
\\
Linear-in-T behavior & \cmark & \xmark      \\
\\
  \hline
\\
Zero-field superconducting diode effect & \xmark & \cmark      \\
\\
    
 \end{tabular}
\end{ruledtabular}
\end{center}
\end{table}

\begin{figure*}[h]
\includegraphics[width=1\linewidth]{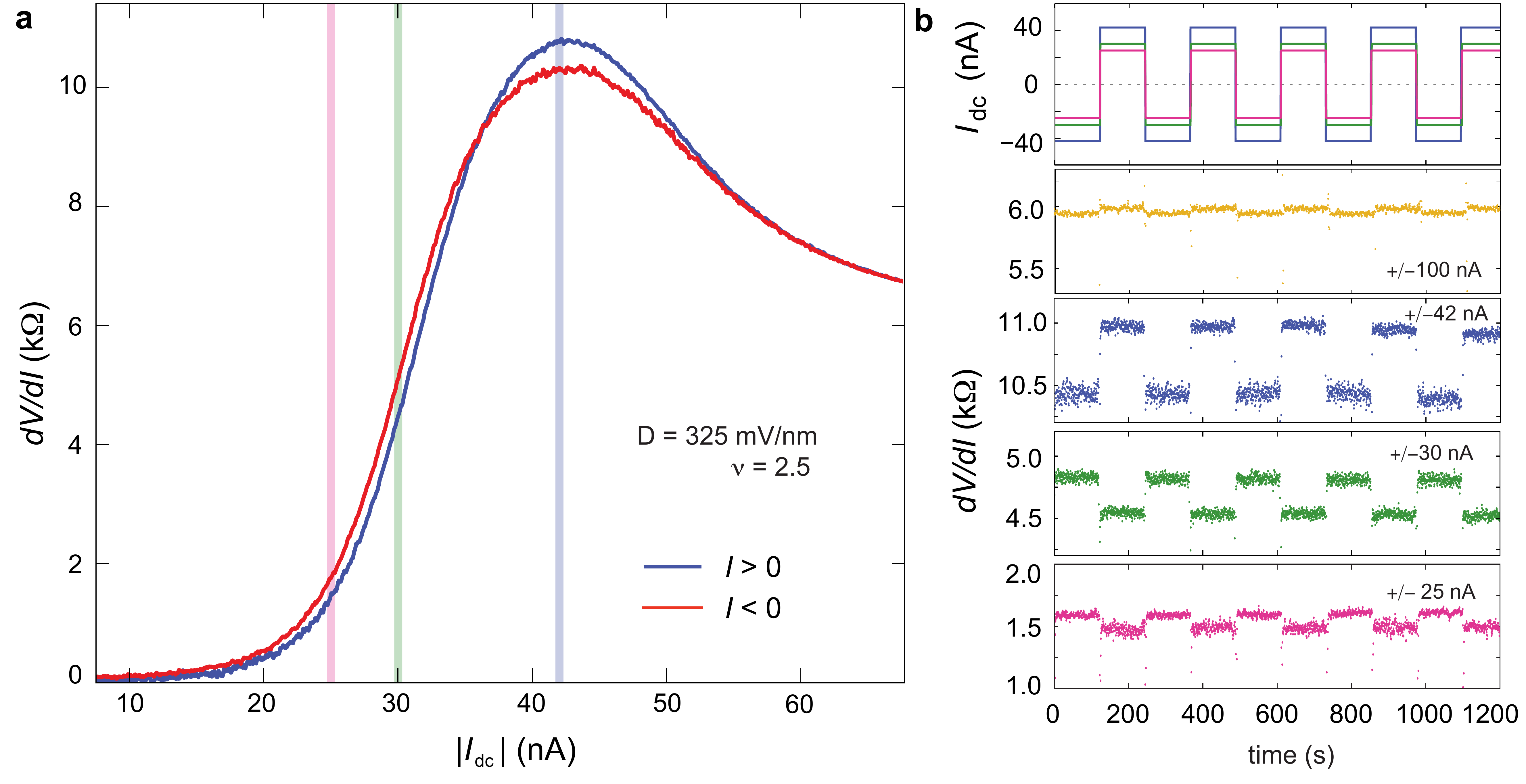}
\caption{\label{fig:Bdiode} {\bf{Zero-field superconducting diode effect in sample B.}} (a) Differential resistance $dV/dI$ as a function of DC current bias $I_{dc}$ measured at $B=0$, $T= 20$ mK and $\nu_{tTLG} = 2.5$.  The blue vertical stripe marks the peak position in the differential resistance, $I_c$, where the superconducting phase transitions into normal state behavior. Blue and red traces denote measurement with positive and negative DC current bias, respectively. The fact that $dV/dI$ measured with different signs of DC current deviate from each other points towards zero-field nonreciprocity in the superconducting transport behavior.  Notably, the nonreciprocity diminishes as DC current exceeds the critical current $I_c$. (b) $dV/dI$ measured with alternating DC current bias at $\pm100$nA, $\pm42$nA, $\pm30$ nA and $\pm25$ nA. Nonreciprocity is apparent in the DC current range of $I_{dc} < I_{c}$. Above the critical current $I_c$, nonreciprocity is substantially suppressed. 
This observation suggests that the fermi surface underlying the superconducting phase is partially valley imbalanced. In this scenario, time-reversal and inversion symmetries are simultaneously broken in the normal state,  and SOC is not required to enable zero-field superconducting diode effect. However, the presence of SOC could still enhance the nonreciprocity through the following mechanisms: (i) SOC enhances the valley polarization in the partially valley imbalanced fermi surface ~\cite{Lin2021SOC}, which gives rise to a more pronounced nonreciprocity in the zero-field superconducting transport behavior; (ii) according to a recent theoretical work \cite{DiodeTheoryPaper},  the presence of the SOC is essential for the trainability of the zero-field superconducting diode effect. Without the SOC-induced trainability, the sample is expected to have multiple domains of opposite valley polarization, diminishing the observed nonreciprocity. These expectations are consistent with our measurement result, where the zero-field diode effect is much weaker in the sample without \WSe. Since two samples do not offer statistical significance to confirm the influence of SOC on the robustness of the zero-field diode effect, we will leave a more systematic discussion on the influence of \WSe\ to a separate work.  It is worth noting that the collection of samples studied in our work is sufficient to confirm the main phenomenology, which is the interplay between orbital ferromagnetism, superconductivity, and Coulomb correlation.
}
\end{figure*}

\newpage
\newpage

\begin{figure*}[h]
\includegraphics[width=1\linewidth]{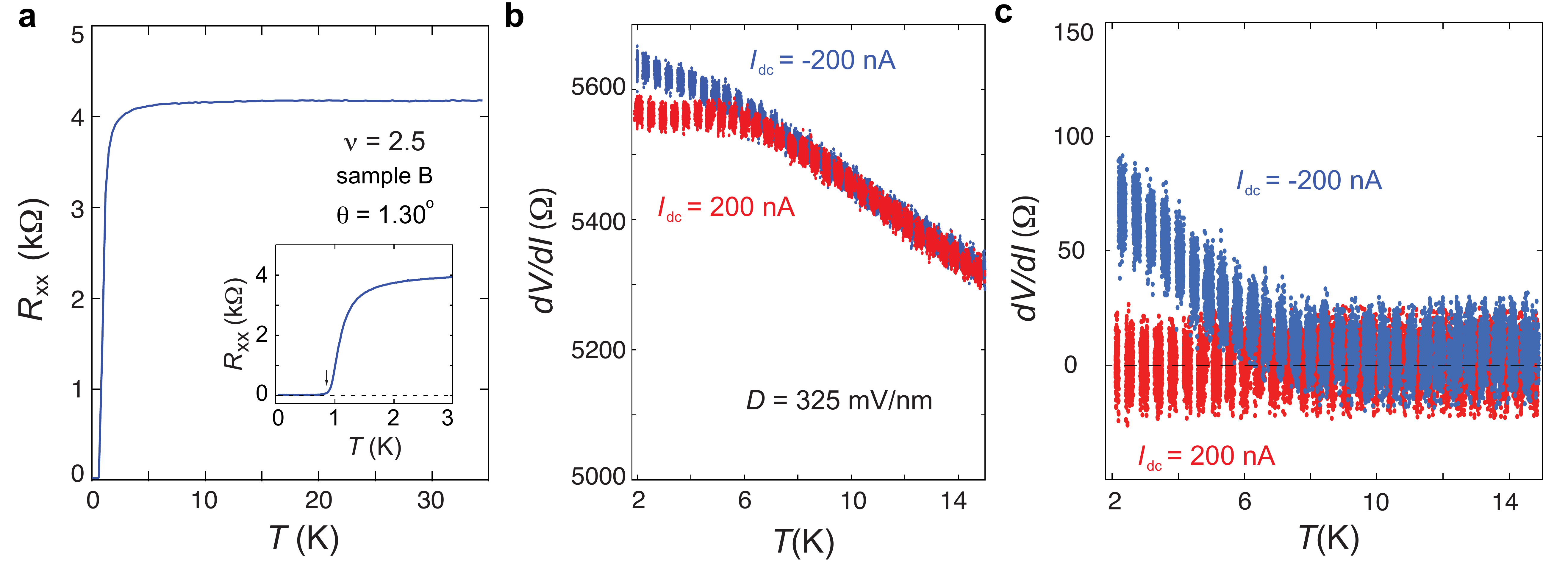}
\caption{\label{fig:Bnormal} {\bf{Weak nonreciprocity in normal state transport in sample B.}} (a) \Rxx\ versus $T$ measured at $B=0$ and $\nu_{tTLG} = 2.5$ in sample B. $T_c$ is around $0.95$ K, defined by the onset in \Rxx\ with increasing temperature. (b) $dV/dI$ measured with $I_{dc} = +200$nA (red dots) and $-200$ nA (blue dots) as a function of temperature in the temperature range $T > T_c$. (c) Renormalized $dV/dI$ showing the difference between $I_{dc} = \pm 200$ nA as a function of $T$. The difference in $dV/dI$ between $I_{dc} = \pm 200$ nA is around $50 \Omega$, which is consistent with the very weak nonreciprocity at $I_{dc} > I_c$ at low temperature (Fig.~\ref{fig:Bdiode}b). This nonreciprocity decreases with increasing temperature and completely vanishes at $T > 6$ K. This observation is consistent with an underlying fermi surface with partial valley imbalance, which simultaneously breaks time-reversal and inversion symmetries. The valley imbalance, hence time-reversal symmetry breaking, onsets  spontaneously at $T \sim 6$ K. Even though time-reversal and inversion symmetries are broken in the normal state, nonreciprocity is greatly enhanced by the onset of the superconducting phase. }
\end{figure*}

\begin{figure*}[h]
\includegraphics[width=0.8\linewidth]{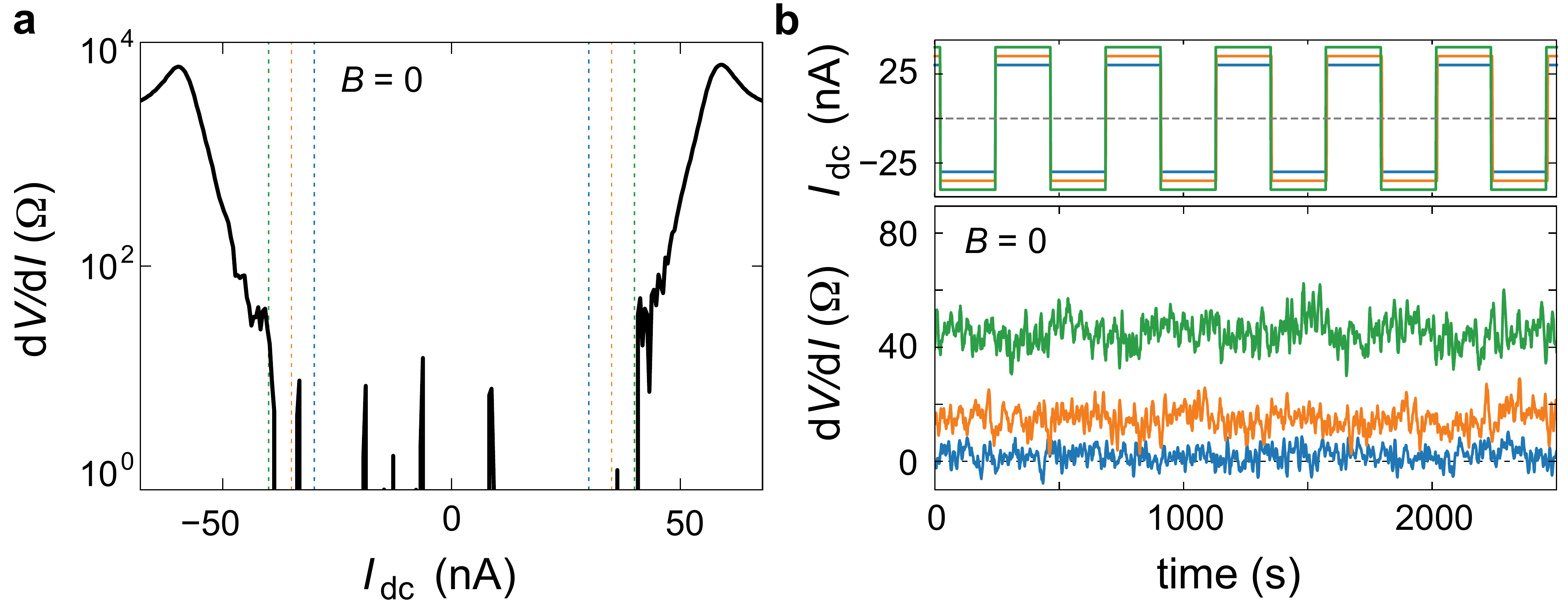}
\caption{\label{fig:Cdiode} {\bf{The absence of zero-field superconducting diode effect near the magic angle.}} (a) Differential resistance $dV/dI$ as a function of DC current bias $I_{dc}$ and an out-of-plane magnetic field $B$ at $T= 20$ mK and $\nu_{tTLG} = -2.64$. Blue marks zero resistance in the chosen color scale. The modulation in $B$ results from Fraunhofer interference. (b) $dV/dI$ as a function of $I_{dc} $ measured at $B=0$. The I-V curve is symmetric with DC bias current. (c) $dV/dI$ as a function of $I_{dc} $ measured at $B=-10$ mT (blue trace) and $+10$ mT (red trace). In the presence of a symmetry breaking field, the IV curve exhibits non-reciprocity, which offers strong indication for proximity induced SOC in sample C. The fact that superconducting diode effect is absent at $B=0$ shows that time reversal symmetry is preserved in both the normal and superconducting phase in sample C. This is distinctly different compared to the observation in sample A. A detailed investigation of sample C is beyond the scope of this current work and we will report this separately. }
\end{figure*}

\newpage

\newpage
\newpage
\clearpage

\subsection{SI 4: Longitudinal and transverse resistance in the normal state} 
We here discuss the connection between the observed non-trivial temperature dependence of \Rxy$/$\Rxx\ and the symmetries of the system. Let $\sigma_{\alpha,\beta}$, $\alpha,\beta=x,y$, be the conductivity tensor, which defines the relation between the in-plane current $I_\alpha$ and electric field $E_{\alpha}$ according to $J_\alpha = \sum_{\beta=x,y} \sigma_{\alpha,\beta} E_{\beta}$. We emphasize that the system has, as opposed to twisted bilayer graphene, no in-plane rotational symmetry (nor any mirror plane perpendicular to the graphene sheets) such that there is no specific in-plane high-symmetry direction that defines a natural orientation of our coordinates $x$ and $y$. We parametrize $\sigma$ as
\begin{equation}
    \sigma = \begin{pmatrix} \sigma_0 & 0 \\ 0 & \sigma_0 \end{pmatrix} + \begin{pmatrix} 0 & \sigma_H \\ -\sigma_H & 0 \end{pmatrix} + \begin{pmatrix} \delta  \sigma_0 & -\sigma_1 \\ -\sigma_1 & -\delta  \sigma_0 \end{pmatrix},
\end{equation}
where $\sigma_H$ is only non-zero if time-reversal symmetry (more precisely $\Theta$ in Table~\ref{ActionOfSymmetries} combined with any spin-rotation) is broken, e.g., as a consequence of the valley-Hall effect discussed in the main text; furthermore, $\sigma_1$ and $\delta \sigma_0$ can only be non-zero if $C_{3z}$ is broken.

Without loss of generality, let us assume that the $x$ axis is chosen along the direction of the applied current in experiment. We then have $R_{xx}=(\sigma^{-1})_{x,x}$ and $R_{xy}=(\sigma^{-1})_{y,x}$. From this it is easy to see that
\begin{equation}
    R_{xy}/R_{xx} = \frac{\sigma_H}{\sigma_0}
\end{equation}
in the case of broken $\Theta$ and preserved $C_{3z}$ (e.g., valley polarization), while
\begin{equation}
    R_{xy}/R_{xx} = \frac{\sigma_1}{\sigma_0-\delta\sigma_0}
\end{equation}
if $C_{3z}$ is broken but $\Theta$ preserved, as is the case in the presence of nematicity. Either way, \Rxy\ is only non-zero if $C_{3z}$ or $\Theta$ (or both) is (are) broken.

One crucial difference between the two scenarios is that \Rxy\ will be isotropic, i.e., invariant under an in-plane rotation of the applied current and measured voltage if it results from broken $\Theta$, while exhibiting a sinusoidal dependence on the rotation angle in the case of a nematicity-driving origin. As such, measuring \Rxy\ or the ratio $R_{xy}/R_{xx}$ as a function of this angle in future experiments will allow to determine its primary origin.

Finally, we comment on the finite and approximately constant value of $R_{xy}/R_{xx}$ at higher temperatures, see in Fig.~\ref{figT}e-f. This can result from admixture of \Rxx\ into \Rxy, e.g., due to a slight misalignment of the voltage probes. It is often modeled \cite{Cao2020nematicity} by an additional artificial contribution to $R_{xy}$ that is proportional to $R_{xx}$, with prefactor that is independent of temperature; this can reproduce precisely the additional temperature-independent part in $R_{xy}/R_{xx}$ that we observe at higher temperatures.

\begin{figure*}
\includegraphics[width=0.65\linewidth]{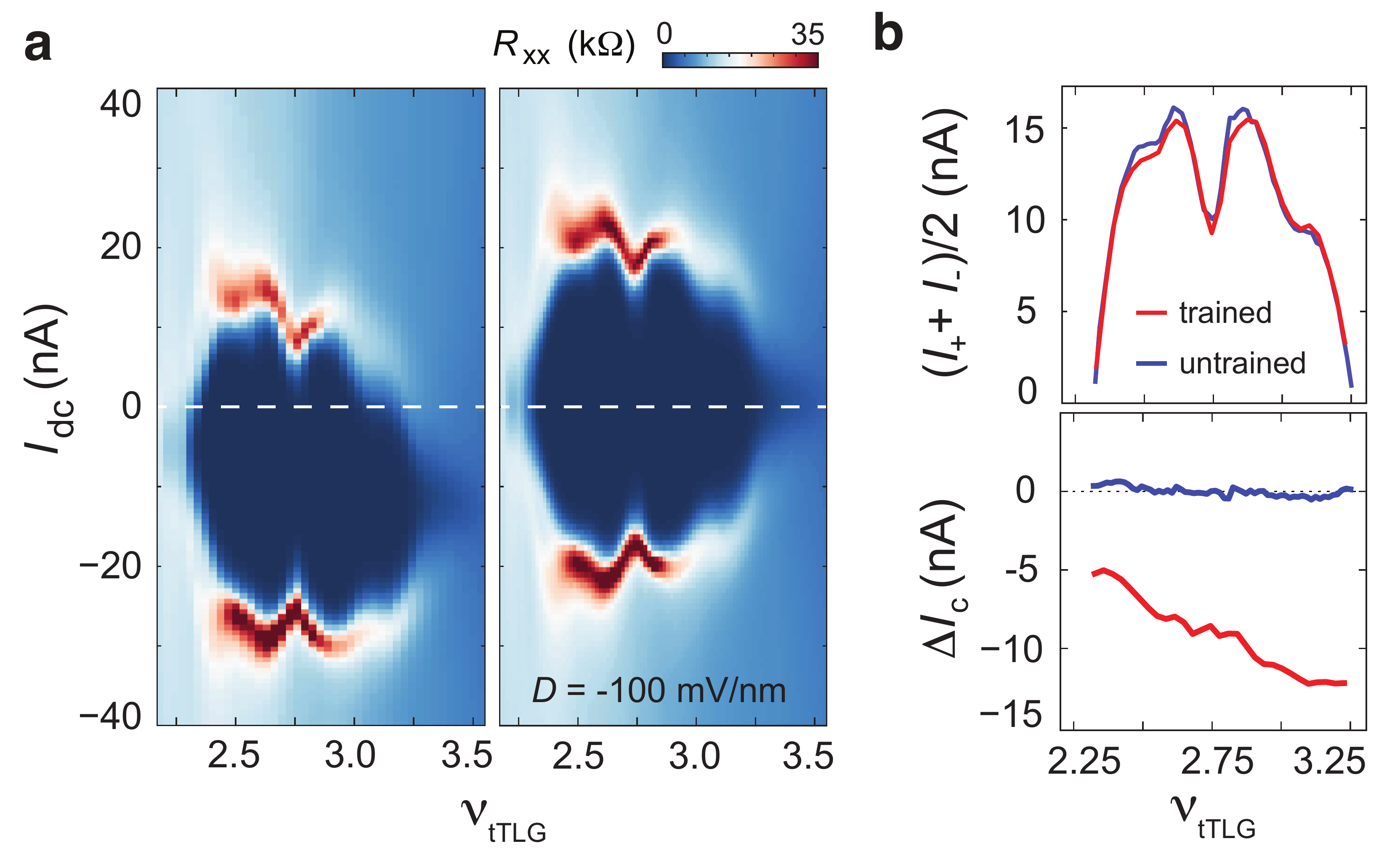}
\caption{\label{fig:Current} {\bf{The robustness of superconductivity in trained and untrained configurations.}} (a) $dV/dI$ as a function of $I_{dc}$ and $\nu_{tTLG}$ measured at $B=0$ and $D=-100$ mV/nm for the electron doped superconducting phase. The measurement is performed with the superconducting diode effect after field training (left panel), and without the superconducting diode effect after ``un-training'' with a large DC current (right panel). (b) The reciprocal (top panel) and  non-reciprocal (bottom panel) component of the critical current, $(I_c^++I_c^-)/2$ and $\Delta I_c$, as a function of $\nu_{tTLG}$ extracted from (a). It is worth noting that several experimental works have reported interplay between DC current flow and the sign of magnetic order: in orbital ferromagnetic states, a large DC current is shown to induce sign-reversal in the magnetic order ~\cite{Serlin2019,Lin2021SOC}. It is hypothesized that the mechanism underlying current-induced switching stems from the interaction between different magnetic domains and current flow around the edge of the domain. Our observation that a large DC current couples to the underlying time-reversal symmetry is consistent with previous experimental results. However, the sample interior of an orbital ferromagnet is insulating and current flows along the edge of the magnetic domain. Whereas the sample interior in the nonreciprocal superconducting phase is highly conductive. As such, we anticipate the interplay between DC current and the underlying time-reversal symmetry breaking to be different. Notably, how DC current interacts with the magnetic order remains an open question for graphene moir\'e systems in general ~\cite{Serlin2019,Lin2021SOC}. In the following, we propose a possible mechanism that could account for the influence of DC current in Fig.~\ref{fig:Currentmodel}.  }
\end{figure*}

\begin{figure*}
\includegraphics[width=0.65\linewidth]{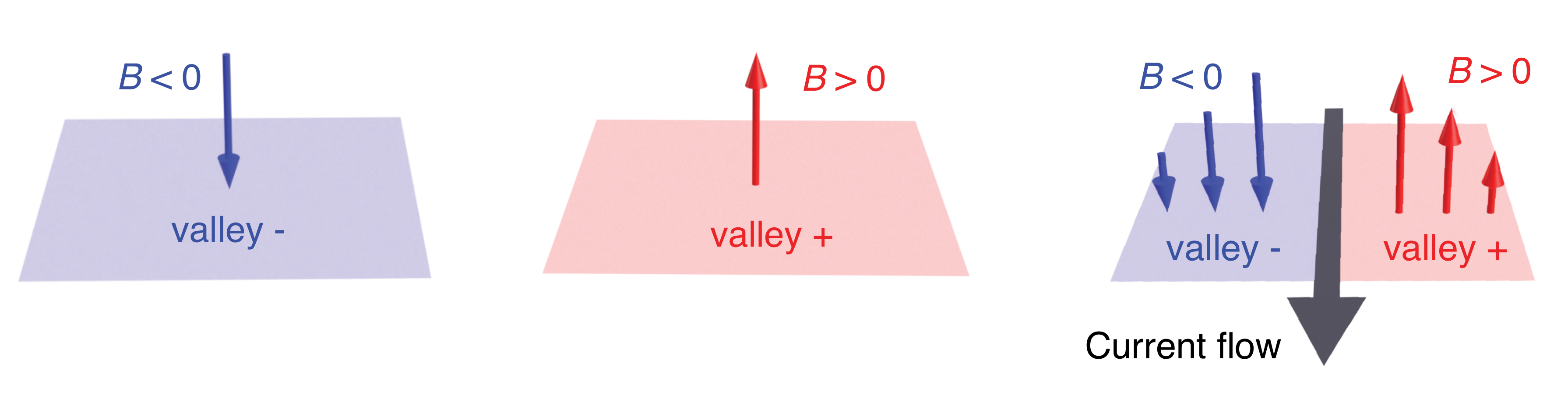}
\caption{\label{fig:Currentmodel} Schematics showing a possible mechanism underlying the influence of a large DC current. As discussed in Ref. \cite{DiodeTheoryPaper}, the sign of partial valley imbalance in the underlying fermi surface can be trained using an external magnetic field in the presence of proximity induced SOC. This trainability provides the basis of our hypothesis that a large DC current untrains the diode effect by generating domains. Graphene moir\'e samples are known to exhibit varying degrees of twist angle inhomogeneity, which could give rise to inhomogeneous distribution of current flow across the sample. In this scenario, a large enough DC current flow could induce a local magnetic field profile which has opposite signs on two sides of the current path. At large DC bias, the current-induced field may be strong enough to stabilize opposite domains. Combined with a spatially inhomogeneous current distribution, this process could, in theory, untrain the superconducting diode effect. }
\end{figure*}

\newpage
\clearpage

\begin{figure*}
\includegraphics[width=0.8\linewidth]{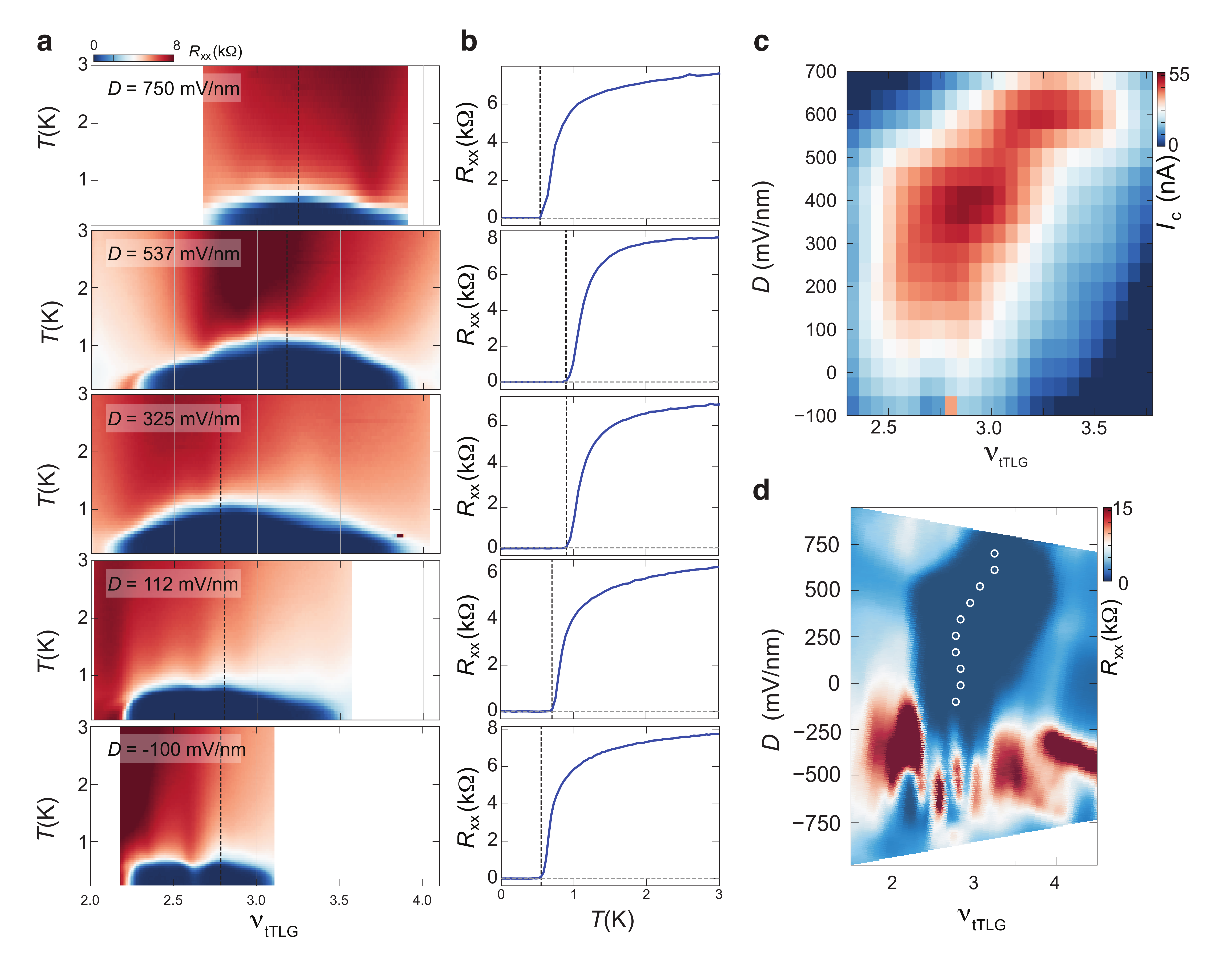}
\caption{\label{fig:optDope} {\bf{Optimal doping and critical temperature dependence on displacement field}}. (a) Longitudinal resistance $R_{xx}$ as a function of filling fraction $\nu_{tTLG}$ and temperature. The optimal doping are indicated via dash lines in (a) and the temperature dependence at optimal doping are plotted in (b). (c) Critical current $I_c$ as a function of filling fraction $\nu_{tTLG}$ and displacement field $D$. The optimal doping for each displacement field is plotted in (d) overlaying with the $R_{xx}$ map. Both the temperature and the dc current dependence suggest the increase in optimal doping as the displacement field increases.}
\end{figure*}

\subsection{SI 5: the influence of proximity effect in sample A} 
The influence of the proximity effect in sample A is reflected by the following observations. The superconducting phase occupies an area in the $n-D$ map which spans the range of displacement field $-800 < D < 800$ mV/nm (Fig.~\ref{figSC}c). According to this map, the superconducting phase exhibits a unique $D$ dependence, which is not symmetric under $D\rightarrow -D$ at fixed $\nu_{tTLG}$, but roughly invariant when the sign of both $D$ and $\nu_{tTLG}$ is flipped. This $D$-dependence is further illustrated by examining the $D$-dependence in the optimal doping of the superconducting phase (Fig.~\ref{fig:optDope}).
The asymmetry in $D$ demonstrates the influence of the \WSe\ crystal,  which is present on only one side of tTLG and, hence, breaks the mirror (and inversion) symmetry of tTLG; the symmetry under $(D,\nu_{tTLG})\rightarrow(-D,-\nu_{tTLG})$ is expected based on the bandstructure of the system ~\cite{DWPaper}. The observed $D$-dependence not only highlights the influence of the tTLG/\WSe\ interface on the band structure but also on the interacting physics, including superconductivity. 
Despite the interfacial effect, \Rxx\ as a function of $T$ exhibits a sharp transition with critical temperature of $T_c \sim 1$ K, as can be seen in Fig.~\ref{figSC}d. The sharp transition is consistent with the excellent sample quality with uniform twist angle.

\subsection{SI 6: Density modulation in sample A} 

The interplay between DW instability and the stability of the superconducting phase has important implications for understanding the tTLG/\WSe\ heterostructure. 
There are two types of interplay, which are observed at  different portions of the phase space: (i) at $D=0$, as shown in Fig.~\ref{figT}a and Fig.~\ref{figDW}a, the stability of the superconducting phase, reflected by quantities such as $T_c$ and $I_c$, exhibits density modulation of $1/4$ moir\'e filling; (ii) at $D = 325$ mV/nm, both $T_c$ and $I_c$ vary smoothly with $\nu_{tTLG}$ without any obvious density modulation. This is true even in the presence of the diverging DOS associated with DW instability. A natural explanation for both of these behaviors is that Cooper pair and DW instabilities compete against each other. As such, a weak superconducting phase exhibts density modulation, whereas a more robust one shows no modulation. We want to draw a clear distinction with the scenario that is discussed in Ref.~\cite{DWPaper}, where the 2 and 4 unit cell states could potentially arise from a 4-fold enlarged moir\'e supercell, which is incuded by twist angle mismatch. In this scenario, superconductivity would be expected to exhibit a density modulation with $1/4$ filling periodicity throughout the phase space. This scenario is inconsistent with our observations.

\begin{figure*}[h]
\includegraphics[width=0.75\linewidth]{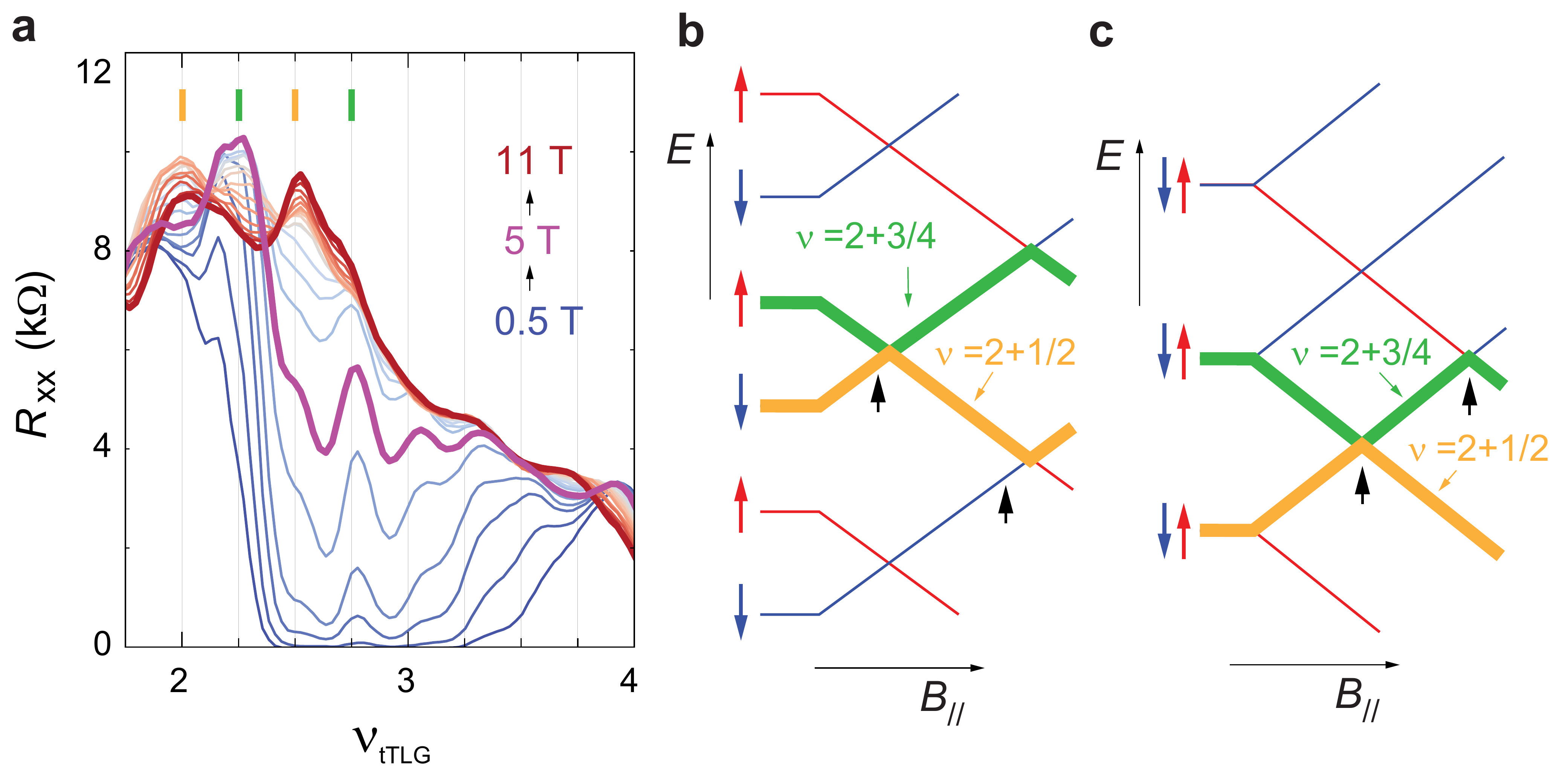}
\caption{\label{fig:Bpara} {\bf{In-plane $B$}} (a) \Rxx\ as a function of $\nu_{tTLG}$ measured at different \Bpara. The vertical red and blue bars mark DW states at every 1/4 fillings which exhibit different B-dependence. (b-c) Band diagram of DW states as a function of in-plane magnetic field. The observation in (a) is consistent with two possibilities: (b) DW sub-bands have alternating spin polarizations at \Bpara\ $=0$; (c) Sub-bands are realigned to be spin degenerate by Cooper pairing instability. The spin degeneracy is lifted by the application of \Bpara.  
Although DW states appear at every 1/4 filling when superconductivity is suppressed by an out-of-plane magnetic field, an in-plane $B$-field gives rise to a doubling in the underlying degeneracy at \Bpara $= 5$ and $11$ T, where DW appears at every 1/2 moir\'e filling, as shown in (a).
At $\nu_{tTLG}=2+1/4$ and $2+3/4$, resistance peaks are robust at  \Bpara\ $\sim 5$ T but become suppressed by a large in-plane Zeeman coupling; on the other hand,
the DW at $\nu_{tTLG}=2$ and $2.5$ are missing around \Bpara $\sim 5$ T appear only at a large enough in-plane magnetic field of \Bpara $\sim 10$ T. 
This unique \Bpara-dependence points towards the possibility that the energy sub-band of DW states features alternating spin polarization. With increasing in-plane Zeeman field, sub-bands with opposite spin index cross at \Bpara $\sim 5$ and $11$ T (b), resulting in a doubling in degeneracy. If the DW states are the parent/normal state for superconductivity, the spin configuration of the superconducting phase is ill-defined. To reconcile these seemingly contradicting observations, the structure of energy sub-bands must be re-arranged at the onset of Cooper pairing instability. As a result, the band structure at $B=0$ and $T=0$ is different from $T>T_c$, or $B>B_c$. One possibility for the re-arranged band structure is shown in (c), where sub-bands with opposite spin polarization become degenerate so that each band is spin unpolarized at \Bpara $=0$. Although it is important to note that other band arrangements are possible as well. }
\end{figure*}

\begin{figure*}
{\includegraphics[width=0.7\textwidth,clip]{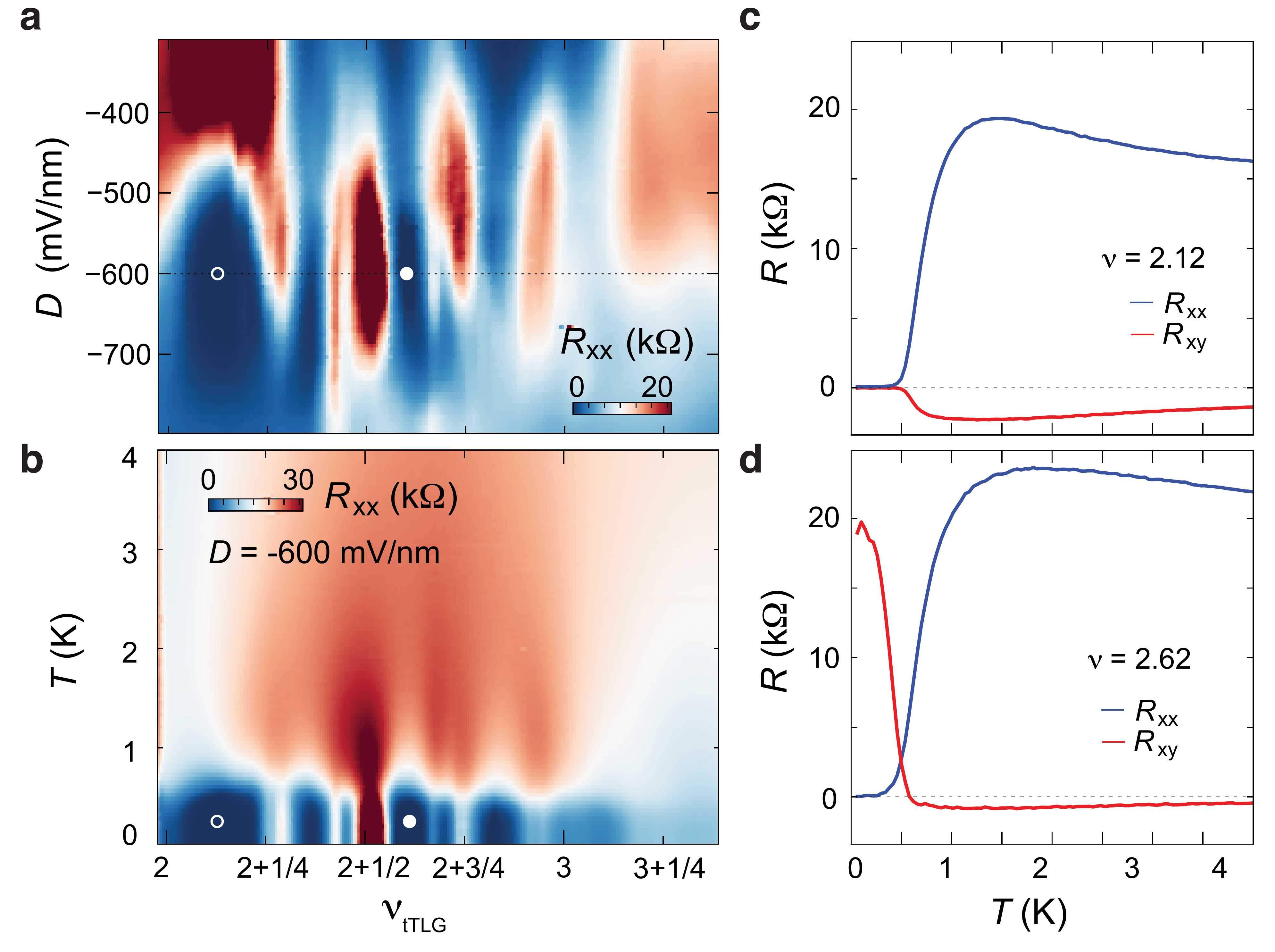}
\caption{\label{f:nematic2}\textbf{Transport anisotropy near $D=-600$ mV/nm.} (a) $\nu_{tTLG}-D$ map of \Rxx\ measured at $B=0$ and $T=20$ mK. In this portion of the phase space, DW and superconductivity coexist at base temperature. (b) $\nu_{tTLG}-T$ map of \Rxx\ measured at $B=0$ and $D=-600$ mV/nm. (c) and (d) Temperature dependence of \Rxx\ and \Rxy\ measured at two positions marked by open and solid white circles in (a). Measurements are performed at $B=0$ and $D=-600$ mV/nm. (c) At $\nu_{tTLG}=2.12$, both \Rxx\ and \Rxy\ diminish at $T < T_c$. (d) At $\nu_{tTLG}=2.62$, as \Rxx\ diminishes with decreasing temperature, \Rxy\ exhibits a sharp onset. At base temperature, the superconducting phase features a large robust Hall resistance. 
As alluded to in the main text, one possible origin of the strongly enhanced \Rxy\ is spatial coexistence with the DW state. Due to the coexistence, superconducting transport naturally inherits the spatial anisotropy of the DW phase~\cite{DWPaper}.
}}
\end{figure*}

\begin{figure*}
\includegraphics[width=0.6\linewidth]{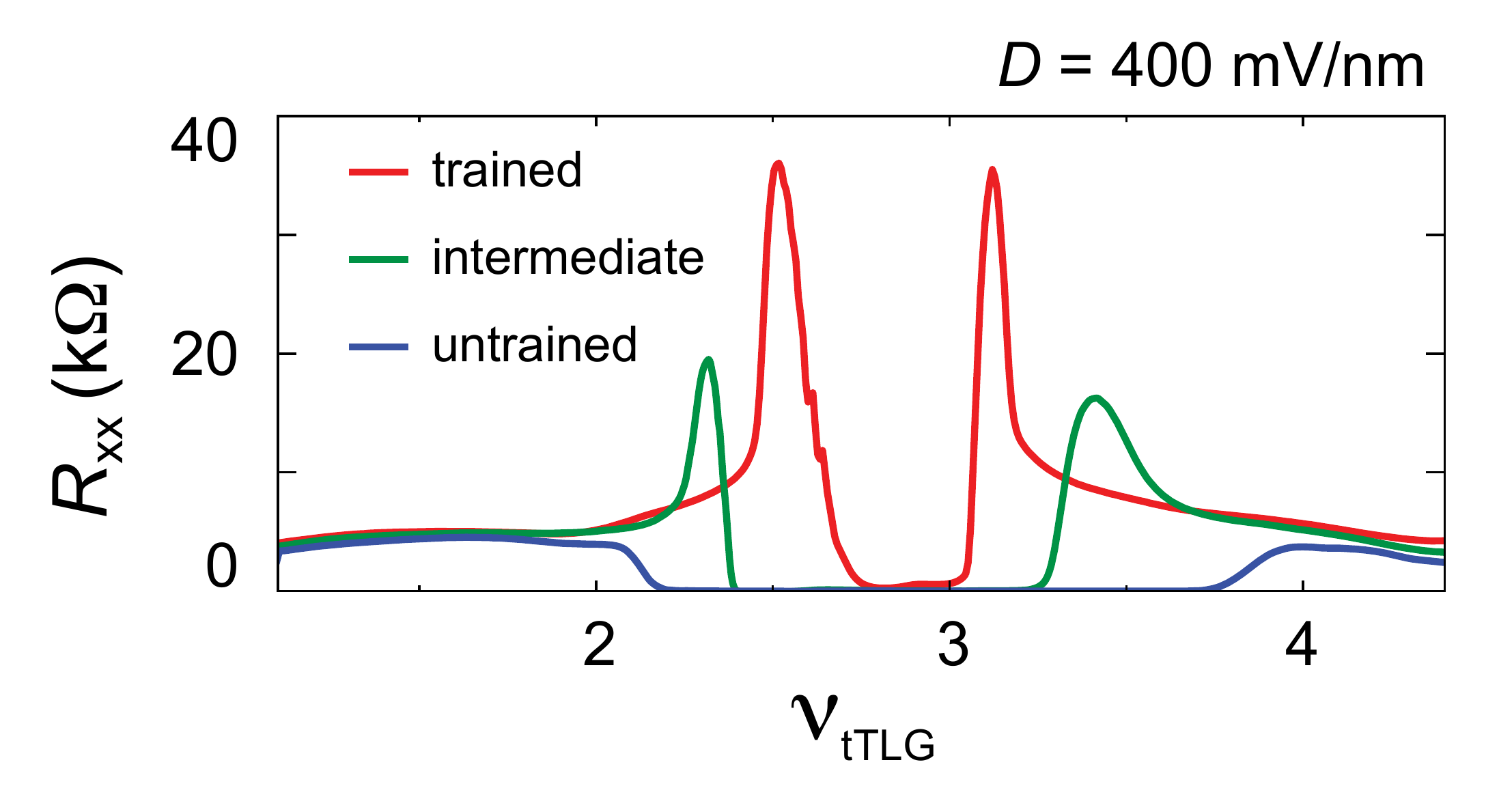}
\caption{\label{fig:Train} \Rxx\ as a function of $\nu_{tTLG}$ measured at $B=0$, $T=20$ mK and $D = 400$ mV/nm. The red trace is measured after field training, where a robust superconducting diode effect is observed. The blue trace is measured after ``untraining'' using a large DC current, where $\Delta I_c \sim 0$. The green trace shows an intermediate configuration, where the nonreciprocity is present but not as robust as the fully ``trained'' configuration.}
\end{figure*}

\begin{figure*}
\includegraphics[width=1\linewidth]{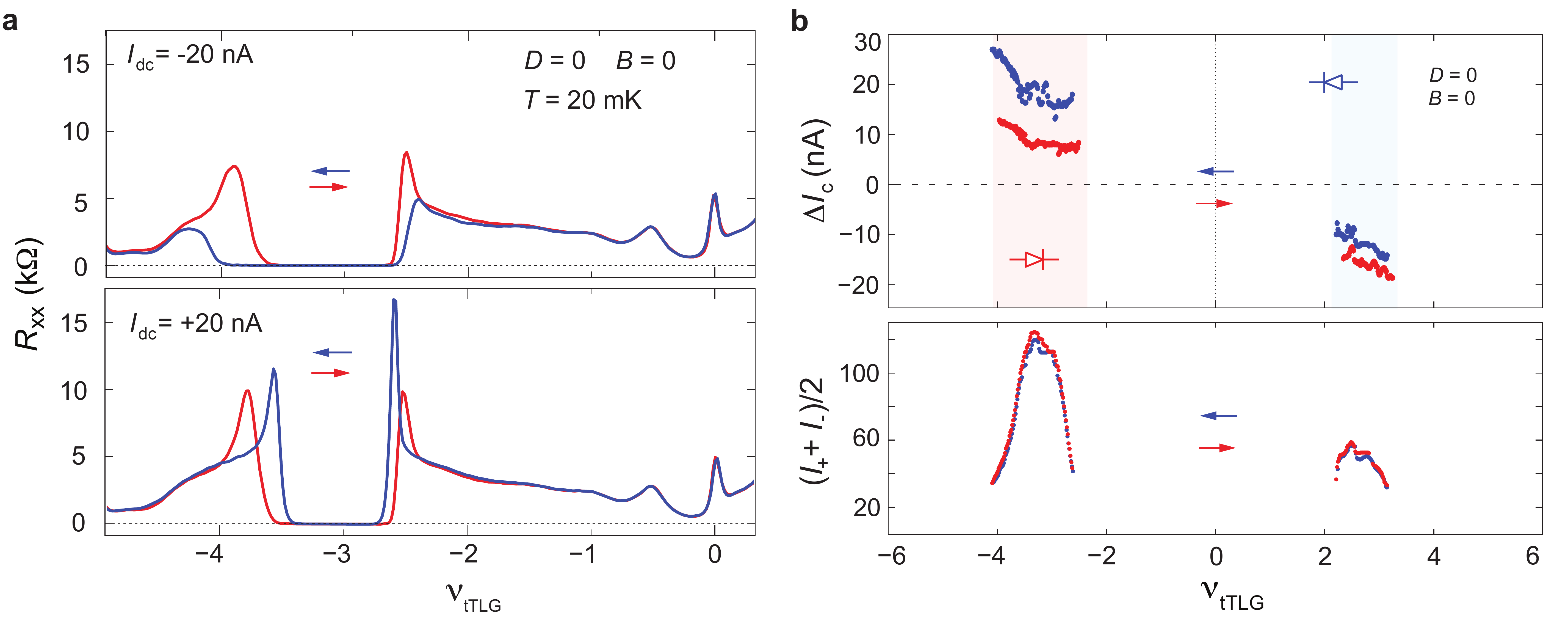}
\caption{\label{fig:hysteresis} {\bf{Hysteresis induced by field-effect doping in sample A.}} (a) \Rxx\ as a function of $\nu_{tTLG}$ measured with $I_{dc} = -20$ nA (top panel) and $+20$ nA (bottom panel). The red and blue traces are measured as carrier density is swept up and down, respectively. (b) $\Delta I_c$ (top panel) and $(I_c^++I_c^-)/2$ as a function of $\nu_{tTLG}$ as the density is swept back and forth. According to (a), the superconducting phase occupies different density ranges depending on the direction of sweeping filed-effect doping. This gives rise to a hysteretic transition between the superconducting-normal phases, which reflects the fact that 
zero-field nonreciprocity is dependent on the direction of sweeping field-effect doping. This is further demonstrated in (b), where the magnitude of $\Delta I_c$ is shown to be dependent on the direction of sweeping carrier density, even though the robustness of the superconductor, defined as $(I_c^++I_c^-)/2$, remains the same.     }
\end{figure*}

\begin{figure*}
\includegraphics[width=0.9\linewidth]{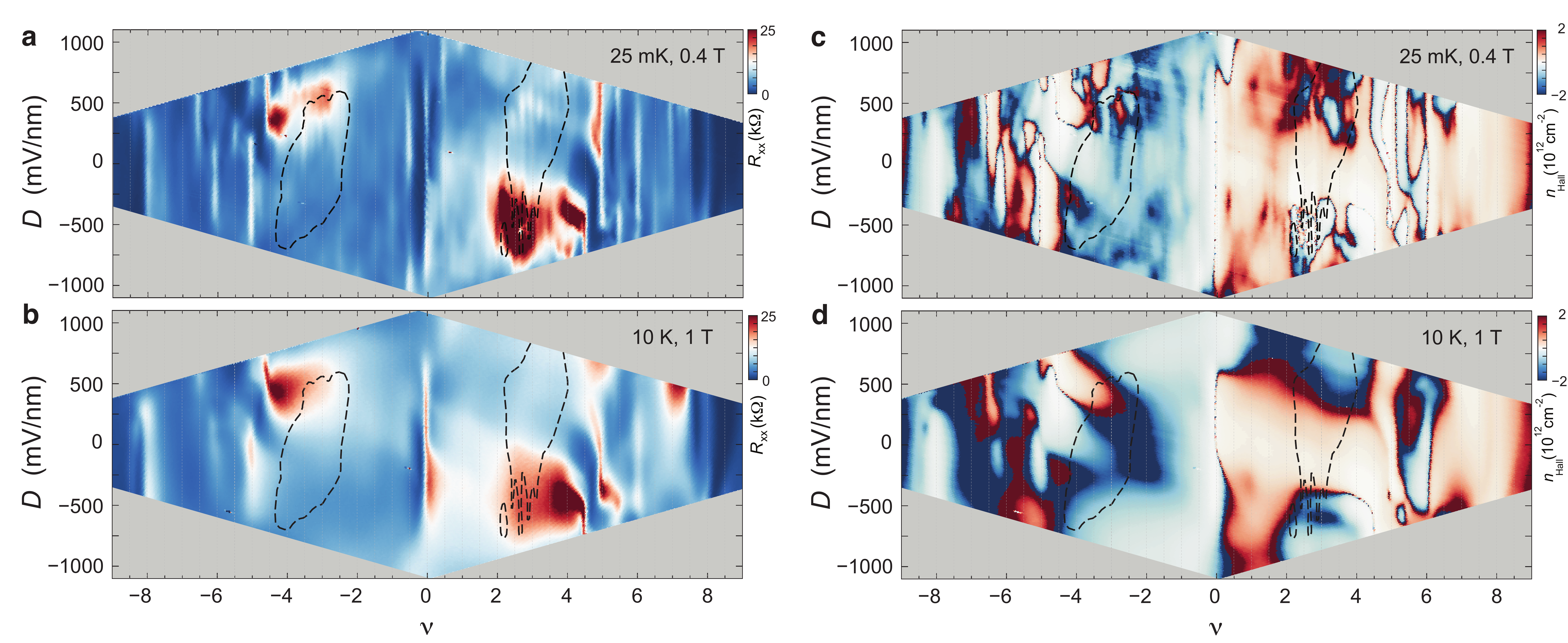}
\caption{\label{fig:10K} {\bf{Longitudinal resistance and hall density at 20 mK and 10 K.}} (a-b) Full range longitudinal resistance $\nu_{tTLG} - D$ map at (a) $T = $ 20 mK, $B = $ 0.4 T and (b) $T = $ 10 K, $B = $ 1 T. (c-d) Full range hall density $\nu_{tTLG} - D$ map at (c) $T = $ 20 mK, $B = $ 0.4 T and (d) $T = $ 10 K, $B = $ 1 T. The dashed lines outline the region of superconductivity.}
\end{figure*}

\begin{figure*}
\includegraphics[width=1\linewidth]{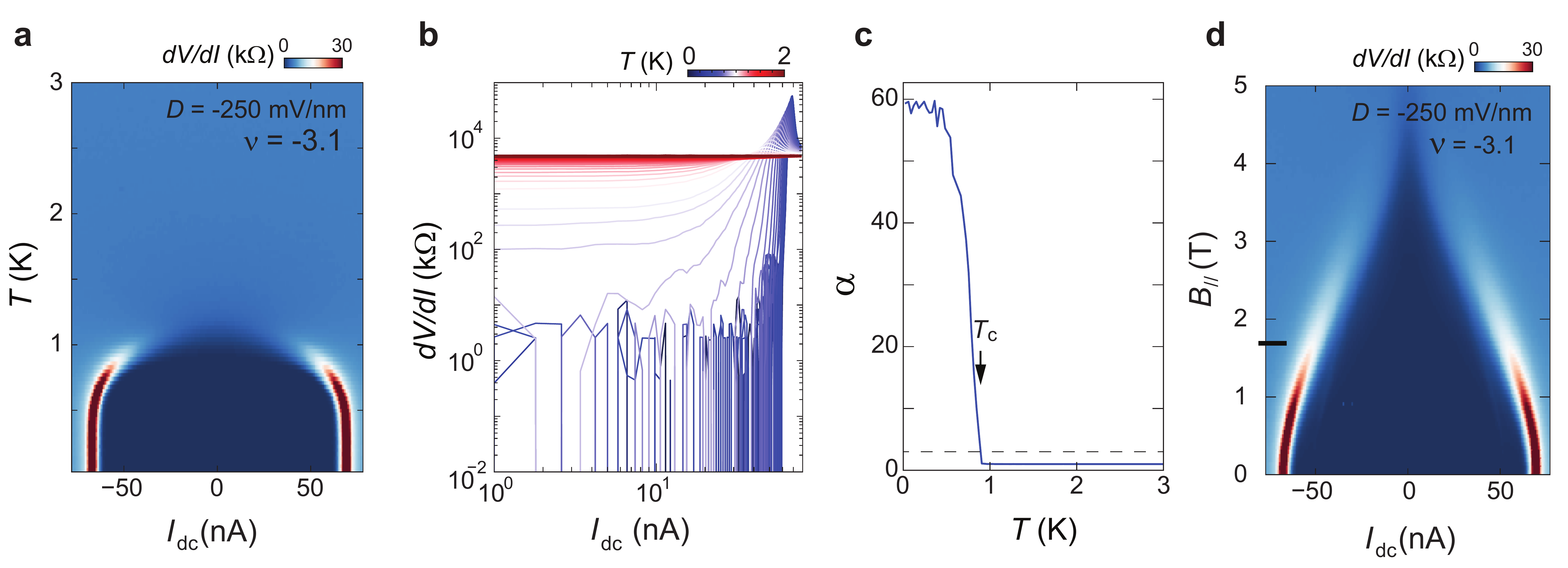}
\caption{\label{fig:Pauli} {\bf{Pauli-limit breaking}} (a) Differential resistance as a function of temperature and current, measured at $D = -250$ mV/nm and $\nu = -3.1$. (b) Log-log plot of differential resistance versus current at different temperatures. Data taken from (a). (c) $\alpha$ as a function of temperature, where $\alpha$ is the exponent in $V = I ^\alpha$. $\alpha$ is extracted by adding one to the slope of the linear region in (b), which is defined by the first ten data points above the noise floor. Critical temperature $T_c = 0.9 K$ is determined as the K-T transition temperature where $\alpha = 3$ (horizontal dashed line). (d) Differential resistance as a function of parallel magnetic field and current, measured at the same $D$ and $\nu$ as (a). The black bar marks the Pauli-limit $B_{//} = 1.86 T/K \times T_c$ for weakly coupled BCS superconductors \cite{Chandrasekhar1962Pauli, Clogston1962Pauli}. Here the superconductivity clearly violates the Pauli-limit as the critical parallel field exceeds the black bar.}
\end{figure*}

\begin{figure*}
\includegraphics[width=0.55\linewidth]{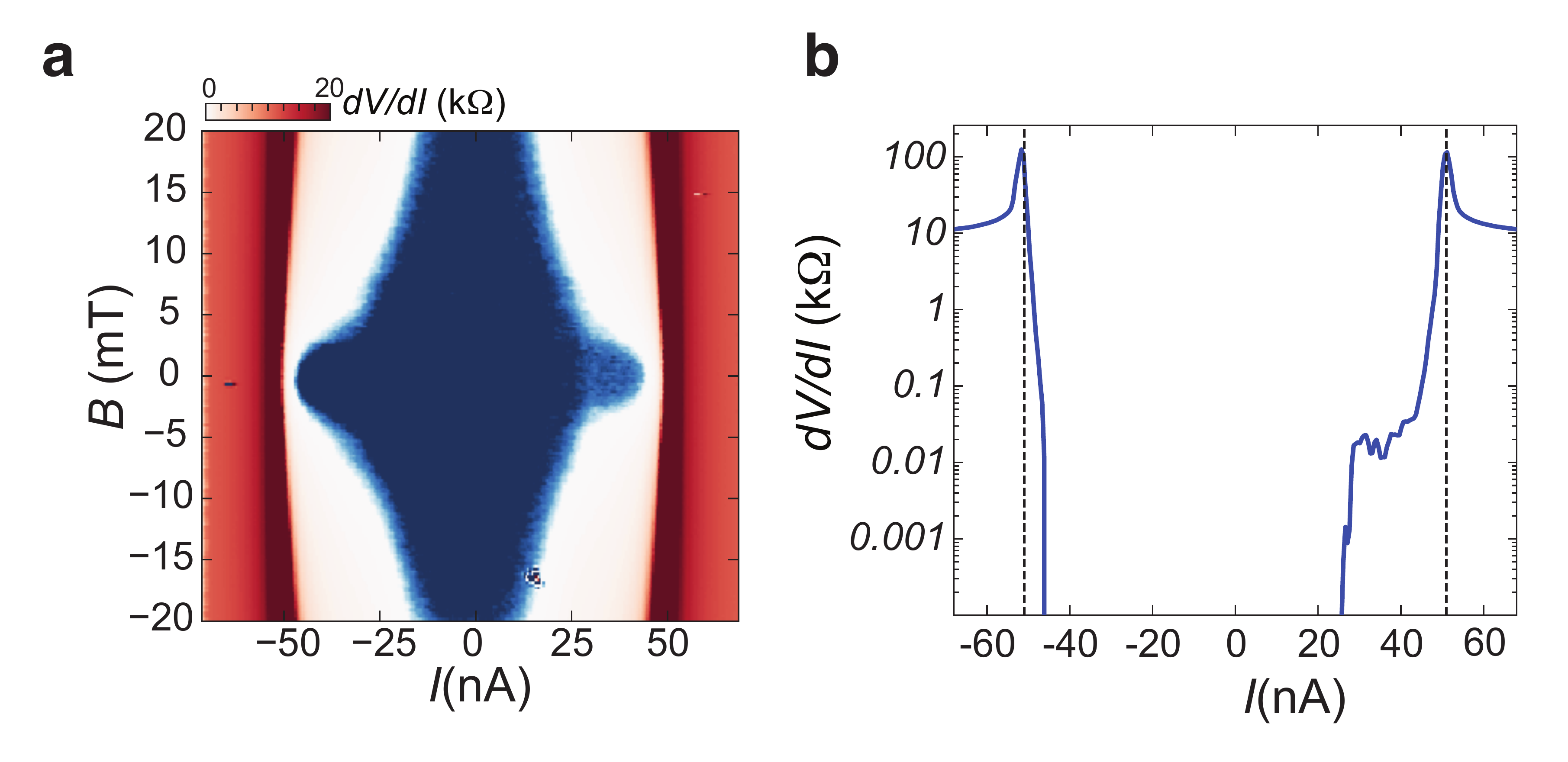}
\caption{\label{fig:IVBasym} {\bf{IV curve from sample A}} (a) Differential resistance $dV/dI$ as a function of the dc current $I_{dc}$ and perpendicular magnetic field $B$ measured at $n_{tTLG} = 2.90 \times 10^{12}$ cm$^{-2}$ and $D = 570$ mV/nm (b)  $dV/dI$ as a function of $I_{dc}$ at $B = 0$ T. Although the peak position in $dV/dI$ is the same for both signs of $I_{dc}$, the onset of $dV/dI$ is highly direction dependent at $B=0$ T. This behavior likely results from magnetic domains in the sample. }
\end{figure*}

\pagebreak

\clearpage

\end{widetext}
\end{document}